\documentclass[aps,pre,preprint,onecolumn,citeautoscript,superscriptaddress,nofootinbib,eqsecnum]{revtex4-1}  
\usepackage{subfig}
\usepackage{graphicx}
\usepackage{amsmath,amssymb,amsthm,amsfonts}
\usepackage{multirow,array,bm,bbm,esint}
\usepackage{epsf}
\usepackage{float}
\usepackage{slashed}
\usepackage[papersize={8.5in,11in}]{geometry}
\usepackage[colorlinks=true]{hyperref}  

\usepackage{color}

\hypersetup{
    bookmarks=true,         
    unicode=false,          
    pdftoolbar=true,        
    pdfmenubar=true,        
    pdffitwindow=false,     
    pdfstartview={FitH},    
    pdftitle={Lefschetz Thimbles and Quantum Phases in Zero-Dimensional Bosonic Models},    
    pdfauthor={R Bharathkumar and Anosh Joseph},     
    pdfsubject={},   
    pdfcreator={Creator},   
    pdfproducer={Producer}, 
    pdfkeywords={} {} {}, 
    pdfnewwindow=true,      
    colorlinks=true,       
    linkcolor=magenta, 
    citecolor=blue,        
    filecolor=magenta,      
    urlcolor=blue           
} 

\def\nn{\nonumber}
\def\bec{\begin{center}}
\def\eec{\end{center}}
\def\beq{\begin{equation}}
\def\eeq{\end{equation}}
\def\bseq{\begin{subequations}}
\def\eseq{\end{subequations}}
\def\bea{\begin{eqnarray}}
\def\eea{\end{eqnarray}}
\usepackage{todonotes}

\begin{document}

\title{Lefschetz Thimbles and Quantum Phases in Zero-Dimensional Bosonic Models}

\author{R Bharathkumar}
\email{r.bharathkumar@outlook.com}

\author{Anosh Joseph}
\email{anoshjoseph@iisermohali.ac.in}

\affiliation{Department of Physical Sciences, Indian Institute of Science Education and Research (IISER) Mohali, Knowledge City, Sector 81, SAS Nagar, Punjab 140306, India}

\date{\today
\\
\vspace{0.8in}}

\begin{abstract}%
\vspace{0.2in}

In this paper, by analyzing the underlying Lefschetz-thimble structure, we investigate quantum phases (or quantum critical points) in zero-dimensional scalar field theories with complex actions. Using first principles, we derive the thimble equations of these models for various values of the coupling parameters. In the thimble decomposition of complex path integrals, determination of the so-called intersection numbers appears as an important ingredient. In this paper, we obtain the analytic expressions for the combined intersection number of thimbles and anti-thimbles of these zero-dimensional theories. We also derive the conditional expressions involving relations among the coupling parameters of the model, that would help us predict quantum phase transitions in these systems. We see that the underlying thimble structure undergoes a drastic change when the system passes through such a phase transition.
 
\end{abstract}

\maketitle

\tableofcontents

\section{Introduction}
\label{sec:Introduction}

We encounter path integrals with complex actions in many branches of physics. The prominent examples are the Minkowski path integral, Yang-Mills theory in the theta vacuum, Chern-Simons gauge theories, chiral gauge theories, and QCD with chemical potential. There are also quantum theories with complex actions that are invariant under ${\cal PT}$ symmetry \cite{Bender:1998ke, Bender:2002vv, Bender:2005tb}. In the context of string theory, the IKKT matrix model, a zero-dimensional supersymmetric quantum field theory that serves as a promising candidate for a nonperturbative formulation of superstring theory, is shown to have a complex fermion operator \cite{Anagnostopoulos:2020xai, Aoki:2019tby, Nishimura:2019qal}. Investigating the nonperturbative structure of such theories using traditional path-integral Monte Carlo methods is unreliable due to the presence of the sign problem. It would be very useful to have a formalism that offers a promising tool to solve quantum field theories containing such complex path-integral weights. 

A recent and developing method to deal with quantum field theories with complex actions uses the complex analog of Morse theory from differential topology \cite{Witten:2010cx, witten_new_2010}\footnote{There exists another compelling method to deal with models containing complex actions. It is based on complex Langevin dynamics. See Refs. \cite{Aarts:2011ax, Aarts:2012ft, Aarts:2013lcm, Aarts:2013uxa, Nishimura:2015pba, Ito:2016efb, Nishimura:2017vav, Basu:2018dtm, Joseph:2019sof, Aoki:2019tby, Nishimura:2019qal, Anagnostopoulos:2020xai} for recent developments in using complex Langevin dynamics in quantum field theories with complex actions.}. There, the objects of primary interest, the so-called Lefschetz thimbles, are a set of sub-manifolds associated with a function that satisfy the Morse flow equation for the real part of the function. The central idea behind using this formalism is to recast the path integral in terms of a finite set of non-oscillatory integrals. Recent work on complex path integrals and connections to Lefschetz thimbles, including applications to quantum tunneling and scattering amplitudes can be seen in Refs. \cite{Guralnik:2007rx, Alexanian:2008kd, Denbleyker:2010sv, Nagao:2011za, Harlow:2011ny, Nishimura:2014rxa, Tanizaki:2014xba, Tanizaki:2014tua, Alexandru:2016san, Ai:2019fri, Ulybyshev:2019fte, Ulybyshev:2019hfm}. In Refs. \cite{Cristoforetti:2012su, Cristoforetti:2014gsa, Cristoforetti:2013wha, Mukherjee:2013aga, Aarts:2013fpa, Fujii:2013sra, Cristoforetti:2014gsa} the Lefschetz-thimble approach has been employed to study bosonic quantum field theories, and in Refs. \cite{Kanazawa:2014qma, Fujii:2015bua, Alexandru:2015xva, DiRenzo:2017igr, Fujimori:2018nvz, Tanizaki:2016cou, Alexandru:2016ejd, Alexandru:2018ngw} models including fermions were studied. The relevance of Lefschetz thimbles in the context of semi-classical expansion in asymptotically free quantum field theories is discussed in Refs. \cite{Dunne:2012ae, Basar:2013eka, Cherman:2014ofa, Cherman:2014xia, Dorigoni:2014hea}.

In this paper, we explore zero-dimensional scalar field theories with complex actions, containing a quartic interaction term and a source term. These models represent the simplest nontrivial quantum field theory with a linear source term. We show that the Lefschetz thimble equations can be derived, using first principles, for various values of the coupling parameters. One result in this paper is the derivation of the expressions of parametrized curves for thimbles and anti-thimbles for all possible cases of the parameters for the quartic model. In the process of thimble decomposition of complex path integrals, an important ingredient is the determination of the so-called intersection numbers. Another result we obtained in this paper is the analytic expressions for the combined intersection number of thimbles and anti-thimbles of these zero-dimensional theories. We also provide a completely analytic demonstration of the existence of quantum phases (or quantum critical points) in the model using the intersection numbers. Due to the lack of existence of a proper definition of thermodynamic quantities in zero dimensions, the discussion is formulated in terms of non-analytic behavior of the partition function. Further, since the locations of these non-analyticities depend only on the (non-thermal) control parameters of the system, a symmetry involving the field $\phi$ is unaffected as one crosses these phase boundaries. These observations indicate that the phases under consideration behave like quantum phases. Conditional expressions involving relations among the parameters of the model help us predict the quantum critical points in these systems. We show that the underlying thimble structure undergoes a drastic change while the system is going through a quantum critical point. Although the accessibility of the information about these phases through Stokes phenomena has been hinted in previous works, see Refs. \cite{Witten:2010cx, Fujii:2015bua, Fukushima:2015qza, Fujii:2015vha}, our work provides the first completely analytic demonstration as a new result.

The paper is organized as follows. In Sec. \ref{sec:A_Primer_on_Lefschetz_Thimbles} we provide a primer on Lefschetz thimbles by introducing the gradient flow equations of the given action. In Sec. \ref{sec:Quartic_Model_with_a_Source_Term} we introduce the model of our interest, a zero-dimensional bosonic model with complex action containing quartic interactions and a source term. The thimble equations for this model are derived next in Sec. \ref{sec:Thimble_Equations_and_Observables}. We discuss analytic expressions for the thimble and anti-thimble equations, and the so-called ghost solutions, which are neither thimbles nor anti-thimbles. We also discuss the behavior of the partition function and observables of the model as a function of the control parameters. In Sec. \ref{sec:Phase_Transition_Boundaries} we discuss the boundaries of phase transitions for various combinations of the values of the coupling parameters. This includes the interesting case when the complex action exhibits ${\cal PT}$ symmetry. A few examples of the phase transition boundaries are provided in Sec. \ref{sec:Phase_Transition_Boundaries_Examples}. The examples show that the structure of the thimbles undergoes a drastic change when the governing (non-thermal) parameters of the model pass through a quantum critical point. In Sec. \ref{sec:Summary_of_Results} we provide a summary of the main results, and in Sec. \ref{sec:Conclusions_and_Future_Directions} we give our conclusions and indicate possible future directions.

\section{A Primer on Lefschetz Thimbles}
\label{sec:A_Primer_on_Lefschetz_Thimbles}

Intuitively, we can relate the Lefschetz thimbles to the original integration cycle of the quantum field theory in the following way. Let us denote the original integration cycle as $\mathcal{M}_{\mathbb{R}}$. We `complexify' this manifold to $\mathcal{M}_{\mathbb{C}}$, that is, we take a complex manifold $\mathcal{M}_{\mathbb{C}}$ that contains the original manifold $\mathcal{M}_{\mathbb{R}}$ as a submanifold, with the requirement that the complex conjugate of an element of $\mathcal{M}_{\mathbb{R}}$ is the element itself. One can think of $\mathcal{M}_{\mathbb{R}}=\mathbb{R}^{n}$ and $\mathcal{M}_{\mathbb{C}} = \mathbb{C}^{n}$ for ease of understanding. 

Post complexification, we identify the \emph{Morse function} \cite{banyaga2013lectures}. The Morse function in a loose sense determines these thimbles. A natural function to consider is the action. (The actual Morse function under consideration is the real part of $-S$ since, by definition, Morse functions are real.) Given a Morse function, we identify its \emph{critical points} -- points in $\mathcal{M}_{\mathbb{C}}$ where the Morse function is locally extremized. The next step, visually, can be thought of as continuously deforming $\mathcal{M}_{\mathbb{R}}$, the deformation being controlled by the Morse function through the Morse flow equations
\beq
\label{eq:morse-flow}
\frac{dz^{i}}{dt} = g^{i\bar{j}} \frac{\partial \overline{S}}{\partial \overline{z^{j}}}, ~~~~\frac{d\overline{z^{i}}}{dt} = g^{i\bar{j}} \frac{\partial S}{\partial z^{j}},
\eeq
where $g^{i\bar{j}}$ is the metric on $\mathcal{M}_{\mathbb{C}}$ and $z_{i}$ are a set of local coordinates around the critical points of $S$. It can be checked immediately that the imaginary part of the action $S$ is constant along the solution to the above equations.

As the final result of this construction, we obtain a pair of sub-manifolds, called the thimble and anti-thimble, associated with each critical point. The thimble is the `stable' solution. That is, the action goes to infinity sufficiently rapidly along a thimble, so as to keep the integral involving $\exp(-S)$ to be convergent. The anti-thimble is the `unstable' solution. An example familiar in physics is the method of steepest descent, and thus the Lefschetz thimbles formalism can be thought of as the generalization of the steepest descent method. A rigorous treatment of this construction can be found in Refs. \cite{banyaga2013lectures, pham1983vanishing, arnol?d2012singularities}.

An integral involving the action on the sub-manifold $\mathcal{M}_{\mathbb{R}}$ can now be written as a linear combination of integrals over the Lefschetz thimbles. In this language, the expression for the partition function associated with a system with action $S$ is given as the weighted sum of contributions from the critical points of the action 
\beq
\label{eq:z-integral}
Z = \sum_i n_i \int_{\mathcal{J}_i} D \phi \; e^{-S[\phi]},
\eeq
where the integral denotes integration over the Lefschetz thimble $\mathcal{J}_i$, which is associated with the $i$-th critical point $\phi_i$ of the action. The weight (also known as the intersection number) $n_i$ is an integer that decides the contribution of a particular critical point to the partition function. Assuming that the critical points do not share a common gradient flow, given in Eq. \eqref{eq:morse-flow}, $n_i$ is given by the number of times the anti-thimble intersects the original integration cycle $\mathcal{M}$ \cite{Tanizaki:2015gpl}. That is,
\beq
\label{eq:int-number-delta}
n_i = \left \langle \mathcal{K}_i, \mathcal{M} \right \rangle.
\eeq

An advantage of using Lefschetz thimbles is that on these thimbles, as discussed above, the imaginary part of the action remains constant. This is certainly a desirable property since, in the (Euclidean) path integral formalism of quantum field theories, the constant imaginary part of the action, ${\rm Im} (S)$, in the integral, Eq. \eqref{eq:z-integral}, can be pulled out as a phase factor, and the remaining integral becomes a non-oscillatory integral\footnote{There is a possibility that the integral can pick up an oscillatory nature due to the Jacobian that transforms the integration measure. This, however, is much milder compared to the original integral and is referred to as the {\it mild sign problem} \cite{Cristoforetti:2012su}.}.

In zero-spacetime dimensions the formalism simplifies greatly. For the majority of the situations considered in this work, the original integration cycle is the real line, $\mathbb{R}$. In this case, we end up dealing with curves in the plane of allowed degrees of freedom for the fields (i.e., $\mathbb{C}$) that satisfy the gradient flow equation
\beq
\label{eq:gradient-flow}
\frac{\partial \phi(t)}{\partial t} = -\overline{\left( \frac{\delta S}{\delta\phi} \right)},
\eeq
where $t$ is a parameter and the overline represents complex conjugation. The thimble ${\cal J}_i$ associated with the critical point $\phi_i$ of the action is defined as the solution to Eq. \eqref{eq:gradient-flow} that satisfies 
\beq
\lim_{t\rightarrow\infty} \phi(t) = \phi_i, \nn
\eeq
and the anti-thimble ${\cal K}_i$ satisfies 
\beq
\lim_{t\rightarrow-\infty} \phi(t) = \phi_i. \nn
\eeq
By definition, the thimbles always end inside regions of stability\footnote{Regions of stability are defined as regions in the complex plane where the integral in Eq. \eqref{eq:z-integral} remains convergent.}, while anti-thimbles end inside regions of instability. 

\section{Quartic Model with a Source Term}
\label{sec:Quartic_Model_with_a_Source_Term}

Let us consider a quantum field theory in zero-spacetime dimensions, with the action given in the following form 
\beq
\label{eq:action}
S[\phi] =  \frac{\sigma}{2} \phi^2 + \frac{\lambda}{4} \phi^4 + h \phi.
\eeq
The action has a quartic interaction term and a source term - it is the simplest nontrivial quantum field theory action with a source term. The  parameters $\sigma$, $\lambda$ and $h$ are in general complex. For convenience, we also express
\beq
\sigma = a + ib ~~ \textrm{and} ~~ \lambda = c + id.
\eeq 

The motivation for considering this particular action is two-fold. First, the above action acts as an excellent toy model for understanding systems with complex actions, in the path integral formalism \cite{Duncan:2012tc, Aarts:2013fpa, Aarts:2013uza, Fukushima:2015qza}, and how Lefschetz thimbles help mitigate the {\it sign problem}, while also being not too trivial and allowing us to showcase a lot of rich dynamics that accompany the Lefschetz thimble analysis. The above action, with complex $\sigma$ is relevant for the relativistic Bose gas at non-zero chemical potential \cite{Aarts:2008wh, Aarts:2009hn}. A variant of this model, with $\sigma = h = 0$ and $\lambda$ complex was studied in Ref. \cite{Duncan:2012tc}.  Second, for the method employed in our calculations, quartic interactions are the highest, exactly solvable terms due to the Abel-Ruffini theorem in algebra \cite{abel_memoire_nodate} that states that there are no closed-form expressions for solutions to general polynomial equations of degree five or higher. Further, the inclusion of a source term ensures that we exhaust all physically possible situations for a system with quartic interactions.

\begin{figure*}[htp!]

\subfloat[$\{a,b,c,d \}=\{0,0,1,-2 \}$]{\includegraphics[width=5cm]{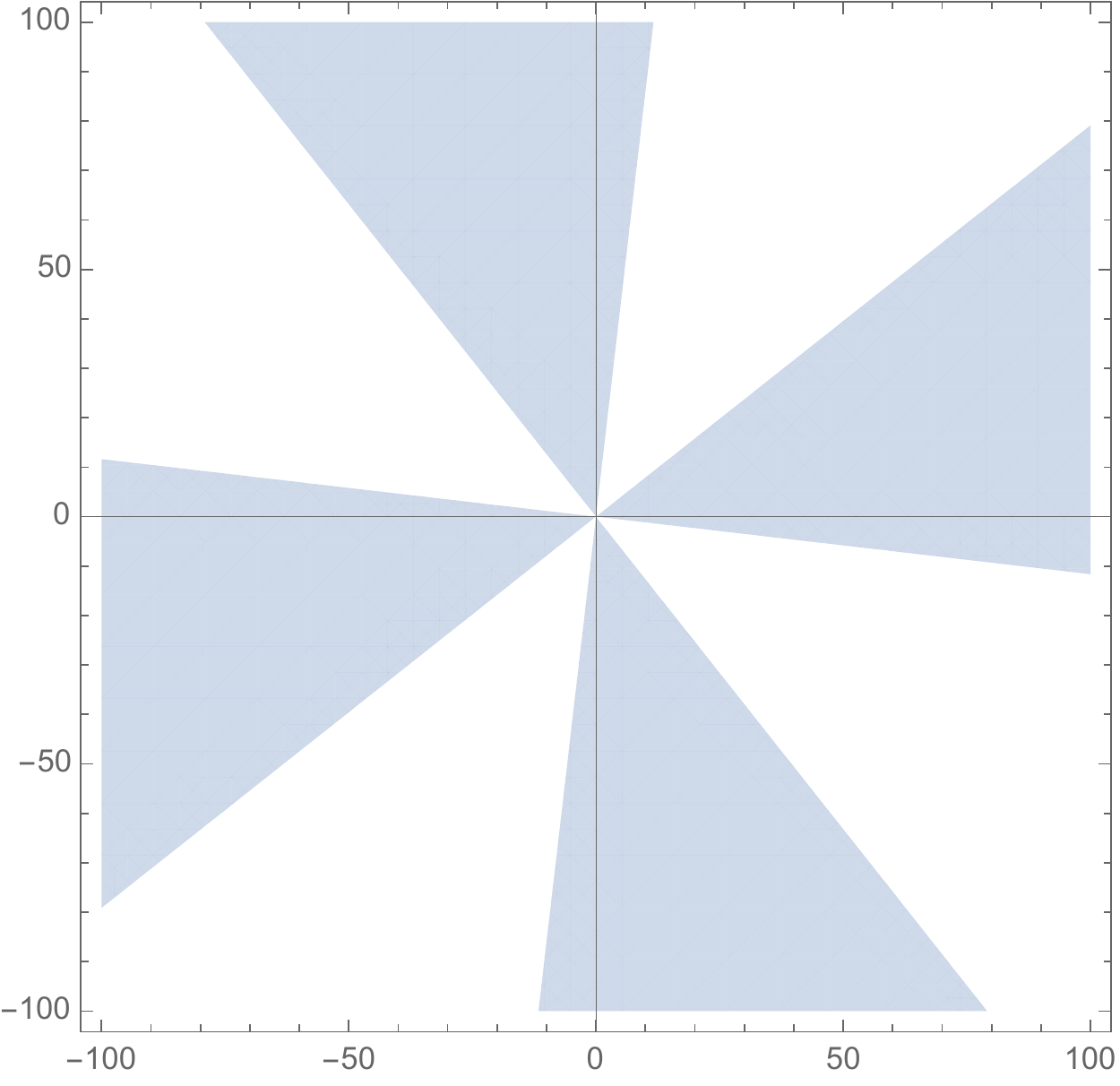}}
$~~~$
\subfloat[$\{a,b,c,d \}=\{0,0,1,0 \}$]{\includegraphics[width=5cm]{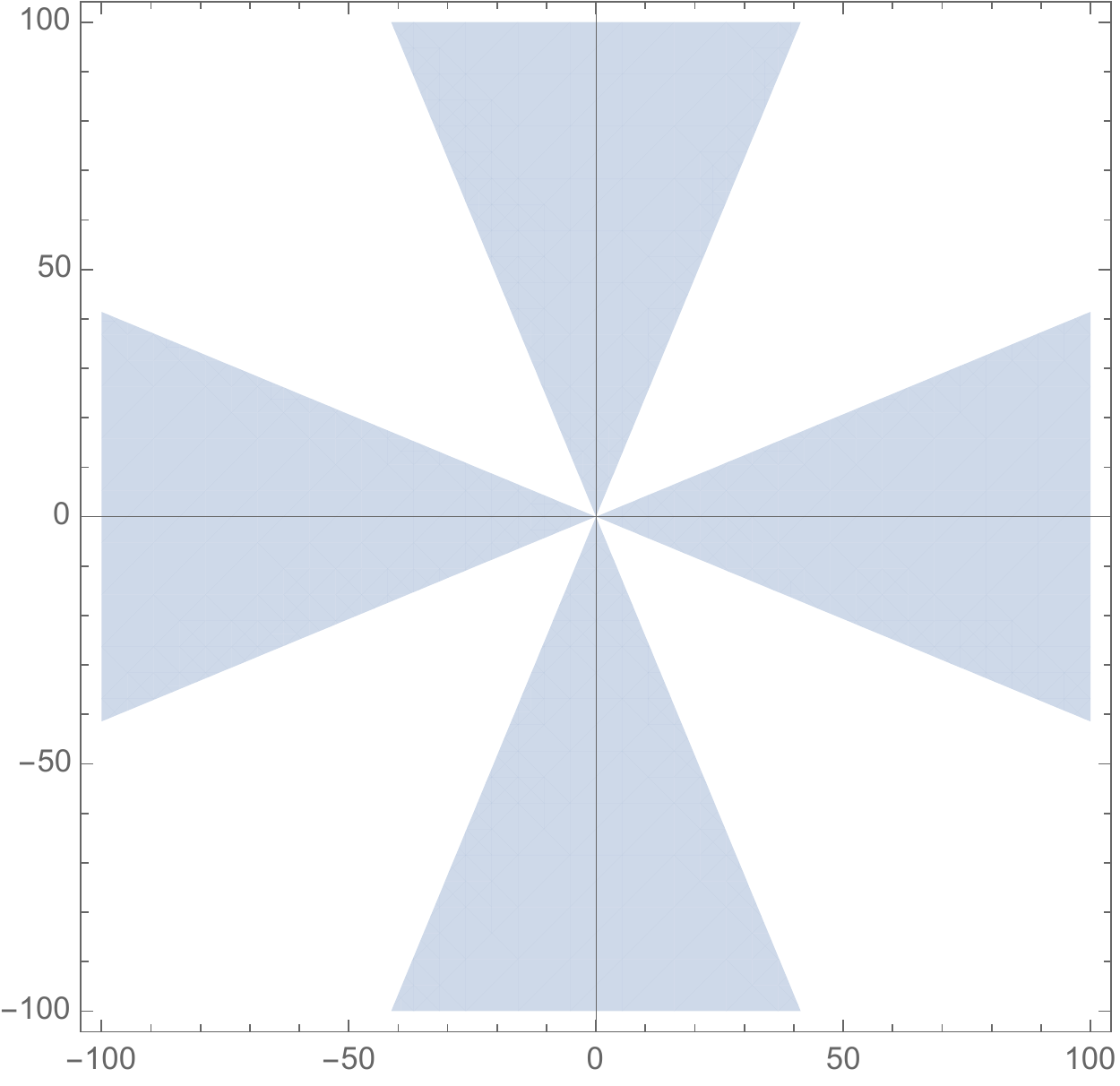}}
$~~~$
\subfloat[$\{a,b,c,d \}=\{0,0,1,2 \}$]{\includegraphics[width=5cm]{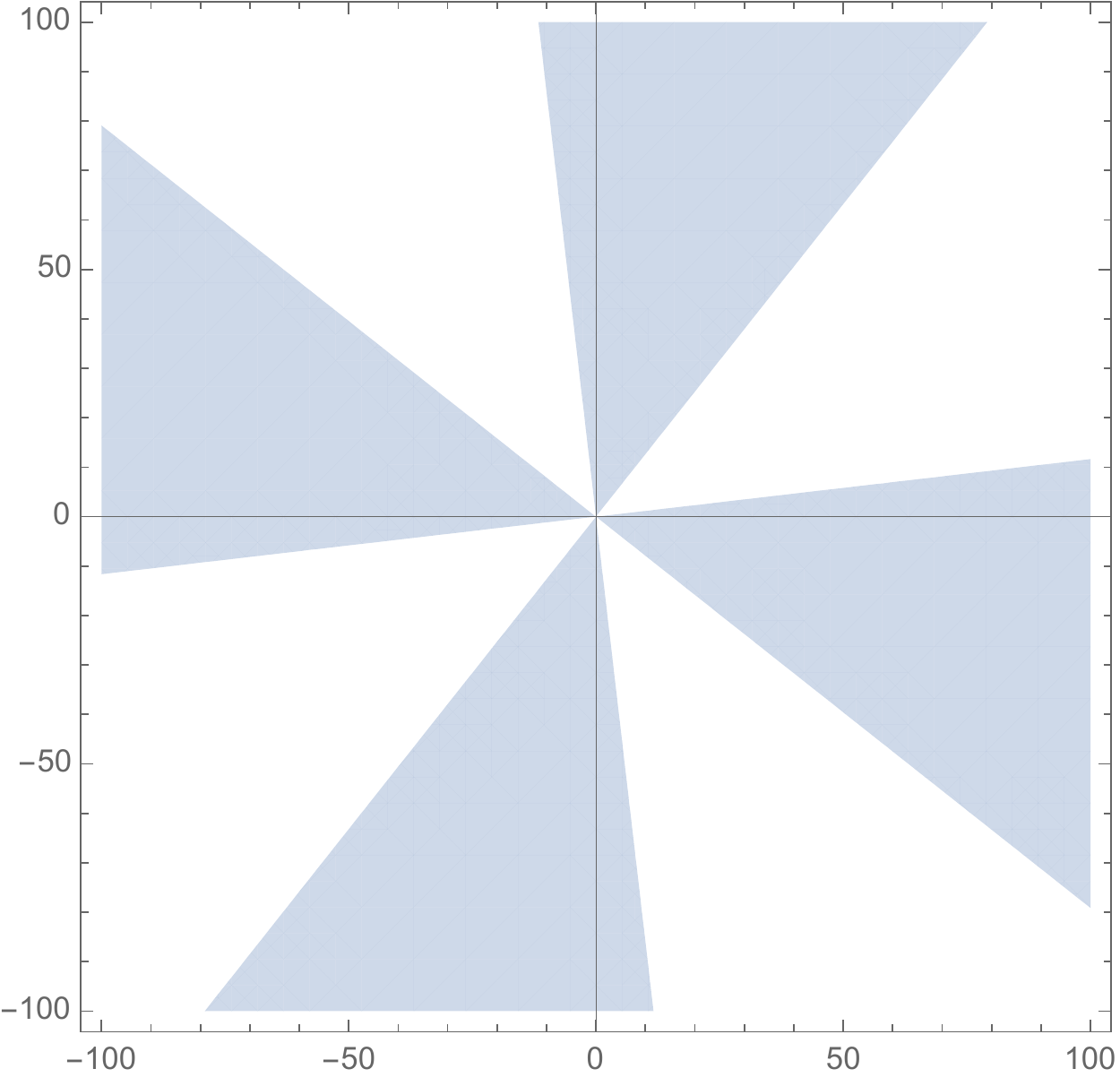}}

\caption{(Color online). A schematic representation of the regions of stability at infinity for the action given in Eq. \eqref{eq:action}. Inside the shaded regions, the integral in Eq. \eqref{eq:z-integral} is convergent. In general, the position and shape of these wedges are controlled by the parameters $\sigma, \lambda$ and $h$ in the action.}

\label{fig:stokes-wedges}

\end{figure*}

Let us begin with determining the regions of stability (sometimes referred to as the {\it Stokes wedges} \cite{Dunne:2015eaa, Cherman:2014ofa}) in this model. Since the integral in Eq. \eqref{eq:z-integral} involves the expression $\text{exp}(-S)$, the integral is convergent in regions where, as $\phi$ approaches infinity, $\text{Re}(S[\phi])\geq0$. Since the highest order in our action is four, we get four wedges in the complex plane where the integral is convergent. This is shown schematically in Fig. \ref{fig:stokes-wedges}. 

One way to find the (anti-)thimble associated with a critical point is to solve the gradient flow equation, Eq. \eqref{eq:gradient-flow}, for (anti-)thimbles. This method, however, quickly becomes very complicated, even for simple forms of actions, due to the coupling between the real and imaginary parts of the differential equation. Fortunately, there is another simpler method. We can exploit a very crucial property of (anti-)thimbles: the imaginary part of the action remains constant along these (anti-)thimbles. Therefore, to solve for the thimbles, we look for solutions to the constraint
\beq
\label{eq:thimble-equation}
\text{Im} \left( S[\phi] - S[\phi_i] \right) = 0,
\eeq
with $\phi_i$ denoting the critical point.

Let us restrict our calculations to cases where $h$ (the parameter controlling the linear term in the action) is small compared to $\sigma$ and $\lambda$. We further restrict $h$ to be either real or purely imaginary. This allows us to approximate the three critical points\footnote{In our discussion here, the critical points are the points in the $\phi$ plane where the action gets extremized, as defined in Sec. \ref{sec:Introduction}. They are not the points in the parameter space corresponding to phase transitions.} of the action as 
\bseq
\label{eq:critical-points}
\bea
\label{eq:criticalpoint1}
\phi_0 &=& -\frac{h}{\sigma} + \mathcal{O}(h^3), \\
\label{eq:criticalpoint2}
\phi_\pm &=& \pm i \sqrt{ \frac{\sigma}{\lambda}} + \frac{h}{2 \sigma}\pm i \frac{3h^2}{8} \sqrt{\frac{\lambda}{\sigma^5}} + \mathcal{O}(h^3).
\eea
\eseq

The critical point $\phi_0$ is close to the origin (that is, $\phi = 0$) for small $h$ while the position of $\phi_\pm$ depends on the choice of the parameters. Let us denote the imaginary part of the action at a given critical point by $\rho_i$. That is,
\beq
\rho_i \equiv {\rm Im} S[\phi_i], ~~i = -, 0, +.
\eeq

For the particular action we are considering, they take the following forms
\begin{widetext}
\bseq 
\label{eq:imag-part-action-i}
\bea 
\rho_0 &=& \left( \frac{b}{a^2 + b^2} \right) \text{Im}(h^2)  + \mathcal{O}(h^3), \\
\rho_\pm &=& \left[ \frac{d(a^2 - b^2) - 2abc}{4(c^2 + d^2)} \right] \pm {\rm Im~} \left( i h \sqrt{\frac{(ac + bd) + i (bc - ad)}{c^2 + d^2}} \right)  \nn \\
&& \hspace{2cm} - \frac{b}{4(a^2 + b^2)} \text{Im}(h^2) + \mathcal{O}(h^3).
\eea
\eseq
\end{widetext}

We note that the convergence of the partition function integral given in Eq. \eqref{eq:z-integral} requires the real part $c$ of $\lambda$ to be positive when the original integration cycle is $\mathbb{R}$. However, when $c$ is negative, which is the case when the action possesses $\mathcal{PT}$ symmetry (we will see this case later), the standard procedure is to take an integration cycle about the angles $5\pi/4$ and $7\pi/4$ in the complex plane (that is, in the third and the fourth quadrant, respectively) \cite{Bender:2007nj, Bender:2006wt}. This choice ensures that the partition function integral remains convergent. If we parametrize the field as $\phi = x + i y$ then this amounts to choosing our integration cycle, in the cases where $c$ is negative, as 
\beq
\label{eq:pt-symmetry-cycle}
y(x) = \left\{
\begin{array}{ll}
      x & ~{\rm for~} x\leq 0, \\
      -x & ~{\rm for~} x > 0. \\
\end{array} 
\right. 
\eeq

\section{Thimble Equations and Observables}
\label{sec:Thimble_Equations_and_Observables}

\subsection{Thimble Equations}
\label{ssec:Thimble_Equations}

As discussed in Sec. \ref{sec:Quartic_Model_with_a_Source_Term}, we solve for the (anti-)thimble by equating the imaginary part of the action at a generic value of $\phi$ to the imaginary part of the action at one of its critical points $\phi_i$. In Ref. \cite{Aarts:2013fpa}, Aarts derived the equation for the (anti-)thimble corresponding to $\phi_0$ when $h = 0$ and $d = 0$. We recreate those results here as a primer, and for completeness. 

Substituting $\rho_0$ into the constraint given in Eq. \eqref{eq:thimble-equation}, having set $h = 0$ and $d = 0$, we obtain the following constraint
\beq
\label{eq:GA-thimble-eq}
-cxy^3 - \frac{b}{2}y^2 + \left( ax + cx^3 \right) y + \frac{b}{2} x^2= 0.
\eeq

Solving for $y$ as a function of $x$, we obtain the thimble and anti-thimble, $\mathcal{J}_0$ and $\mathcal{K}_0$, respectively, associated with the critical point $\phi_0$ 
\bseq
\label{eq:GA-thimble}
\bea
y(x) &=& \frac{1}{6cx} \left(-b + e^{-i\theta}\frac{\Delta_2}{\Delta_1} + e^{i\theta}\Delta_1\right), \label{eq:GA-thimble-y} \\
\Delta_1 &=& \left( b \Delta_3 + \sqrt{b^2 \Delta_3^2 - \Delta_2^3} \right)^{1/3}, \label{eq:GA-thimble-d1} \\
\Delta_2 &=& b^2 + 12 c x^2 \left( a + cx^2 \right), \label{eq:GA-thimble-d2} \\
\Delta_3 &=& b^2 + 18 c x^2 \left( a - 2cx^2 \right). \label{eq:GA-thimble-d3}
\eea
\eseq

\begin{figure*}[t]

\subfloat[Thimbles, anti-thimbles and ghosts for all the critical points, $\phi_0$ and $\phi_\pm$, of the action.]{\includegraphics[width=5cm]{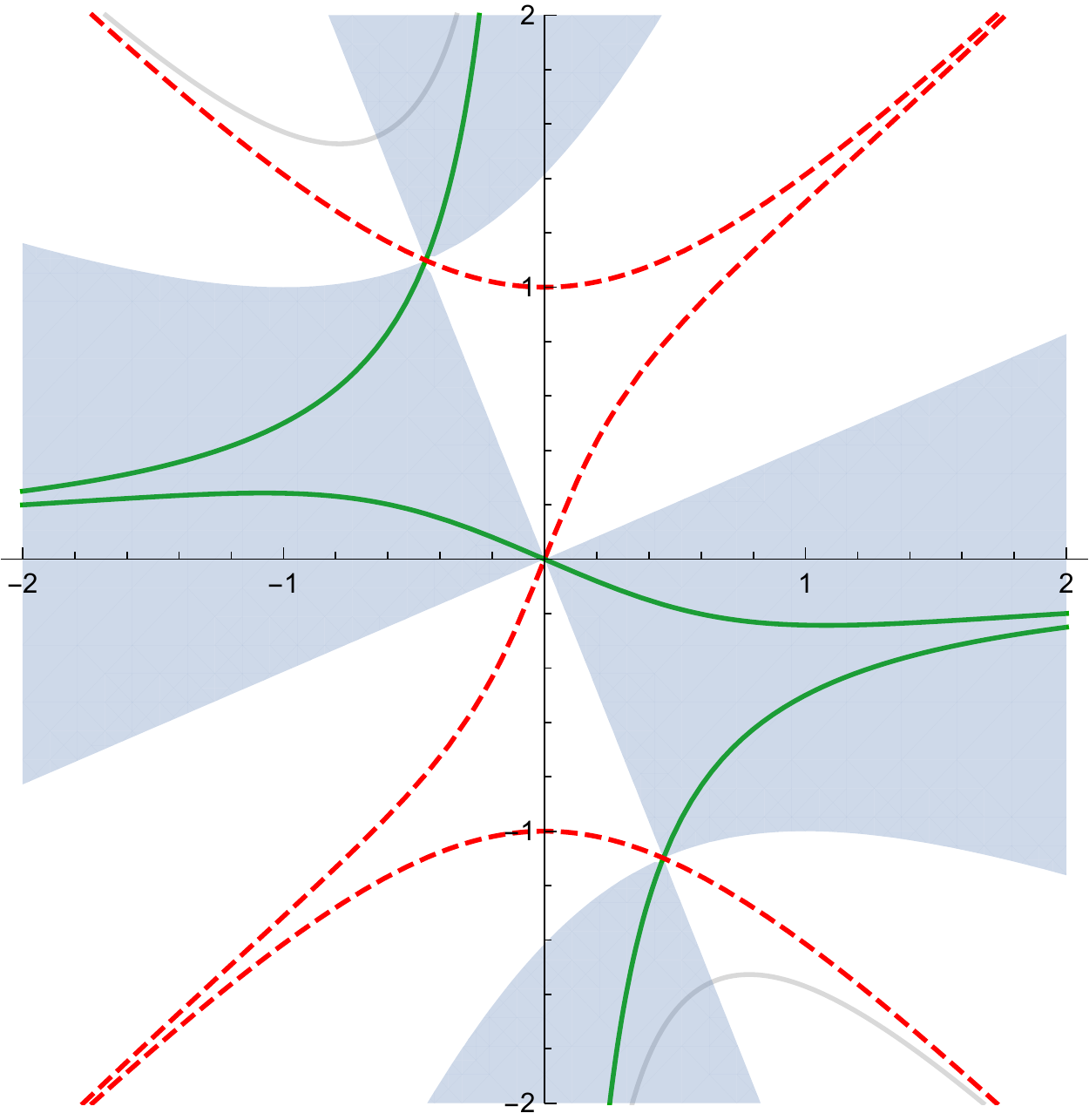}}
$~~~$
\subfloat[Thimble, anti-thimble, and ghosts for the critical point $\phi_0$ of the action.]{\includegraphics[width=5cm]{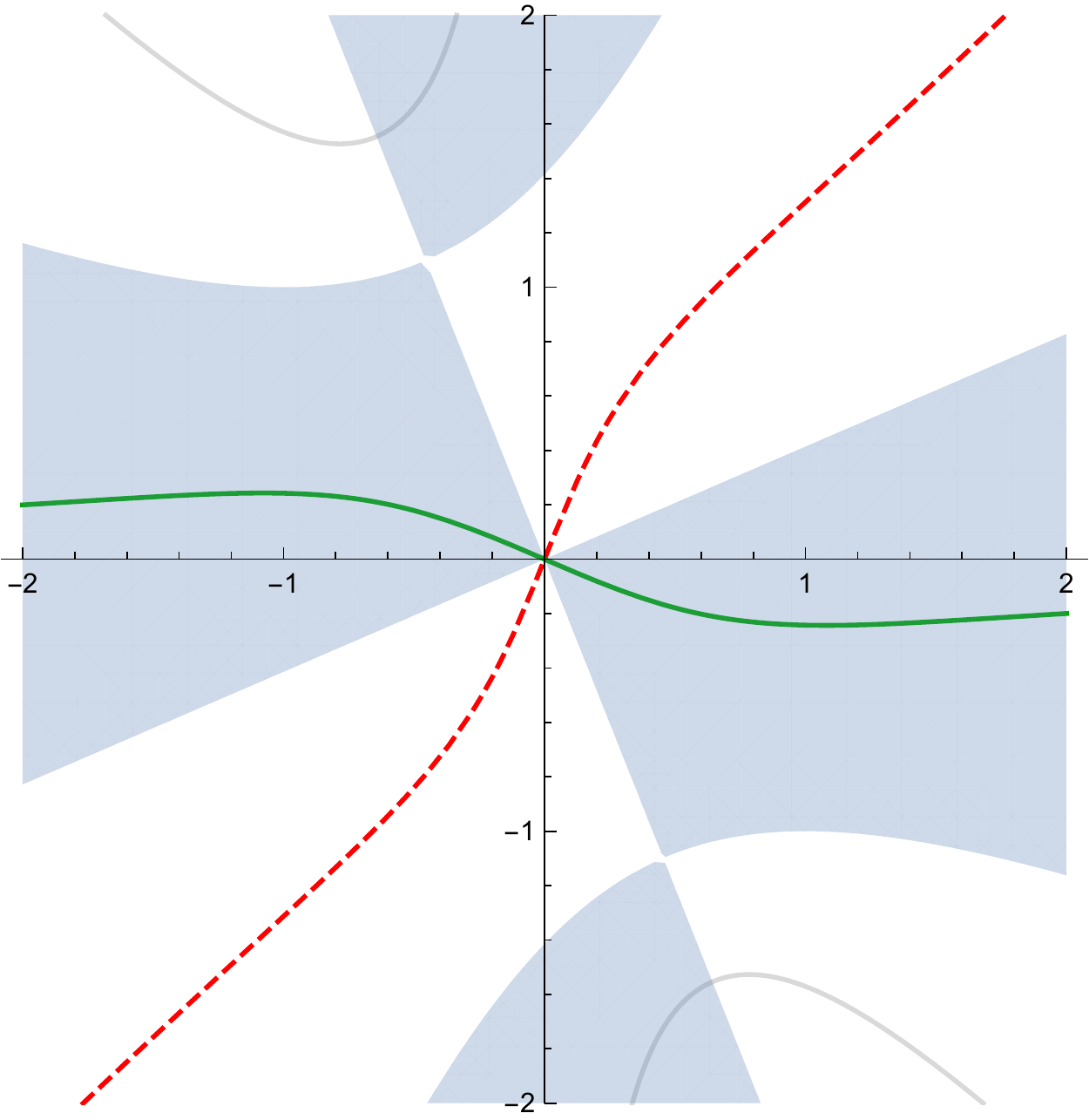}}
$~~~$
\subfloat[Thimbles and anti-thimbles for the critical points $\phi_\pm$ of the action.]{\includegraphics[width=5cm]{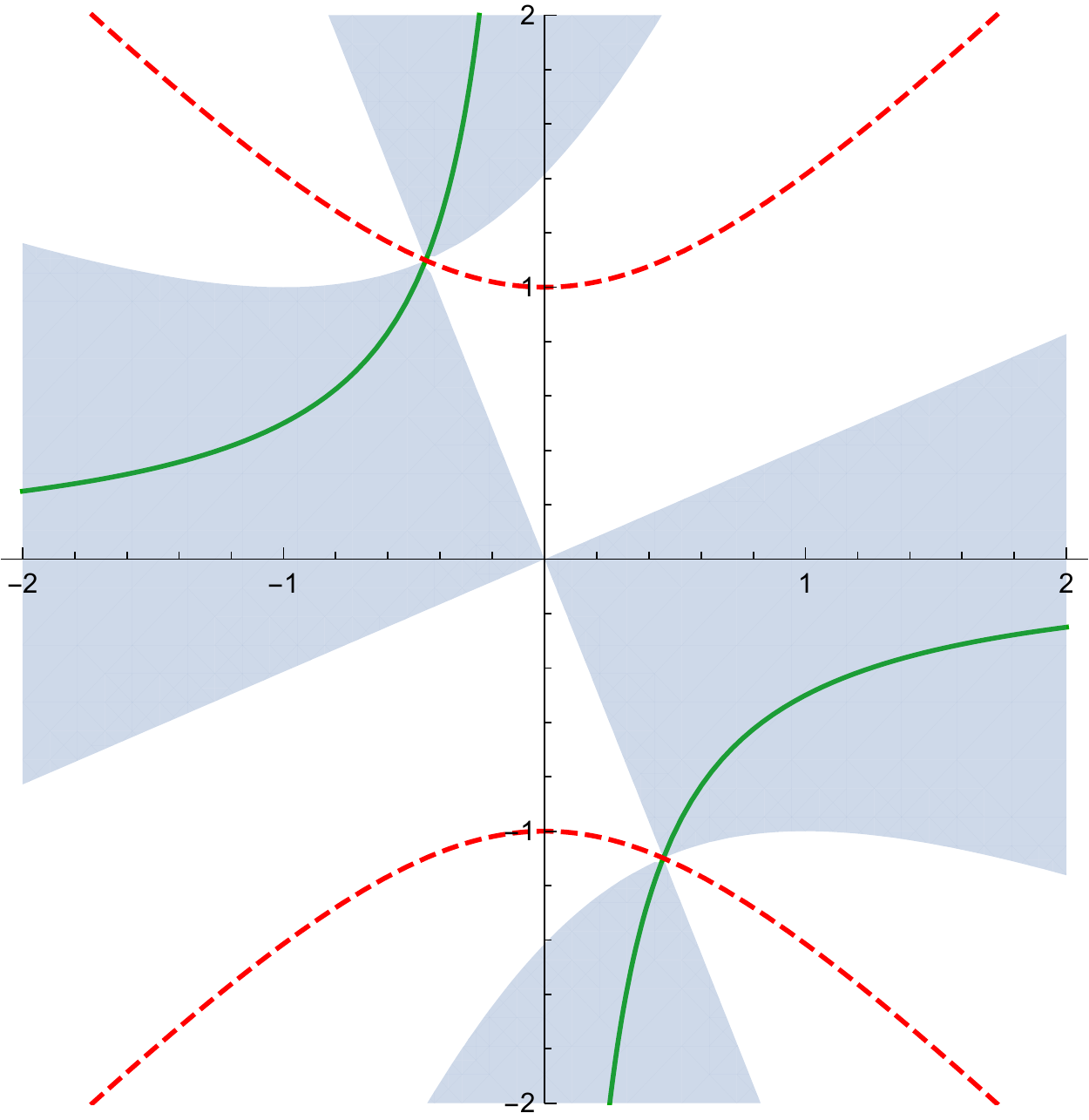}}

\caption{(Color online). The solutions to the thimble equation given in Eq. \eqref{eq:thimble-equation}, corresponding to the critical points $\phi_0$ and $\phi_\pm$, for the parameters $\{ a = 1, b = 1, c = 1, d = 0, h = 0 \}$. In all the three figures, the green solid curves represent the thimbles, red dashed curves represent the anti-thimbles, and the grey solid curves represent the ghosts. The shaded regions represent the regions where $\text{Re}(S)\geq0$.}
\label{fig:GA-thimbles}

\end{figure*}

Here $\theta \in \{ -\frac{\pi}{3}, 0, \frac{\pi}{3} \}$. In Fig. \ref{fig:GA-thimbles} we show the three curves corresponding to the three values of the parameter $\theta$. The thimble corresponds to $\theta = -\frac{\pi}{3}$ and the anti-thimble corresponds to $\theta = \frac{\pi}{3}$. The curves for $\theta = 0$ are paths of constant $\text{Im }S$ that are neither thimbles nor anti-thimbles. We shall refer to these curves as the {\it ghost solutions} or simply {\it ghosts}. 

Similarly, when solving for $\phi_\pm$, we obtain Eq. \eqref{eq:GA-thimble}, but with Eq. \eqref{eq:GA-thimble-d3} now changed to the following form
\beq
\Delta_3 = b^2 + 72 c x^2 \left( a - cx^2 \right).
\eeq

In this case, $\theta = 0$ corresponds to the thimbles for both $\phi_+$ and $\phi_-$. The curve has two branches, one for $x < 0$ and the other for $x > 0$. The anti-thimble associated with $\phi_+$ has $\theta = -\frac{\pi}{3}$. The anti-thimble associated with $\phi_-$ has $\theta = \frac{\pi}{3}$. In Fig. \ref{fig:GA-thimbles} we show the thimbles, anti-thimbles and ghosts for all the critical points, $\phi_0$ and $\phi_\pm$, of the action for the parameters $\{ a = 1, b = 1, c = 1, d = 0, h = 0 \}$.

So far we have restricted the model to the case where $h = 0$ and $d = 0$. Let us now do away with the restriction on $d$ while still maintaining the constraint $h = 0$. The thimble equation given in Eq. \eqref{eq:GA-thimble-eq} is now modified as
\beq
\label{eq:GA-thimble-eq-dneq0}
\frac{d}{4} y^4 - c x y^3 - \left( \frac{b}{2} + \frac{3d}{2} x^2 \right) y^2 + \left( ax + cx^3 \right) y + \frac{b}{2} x^2 + \frac{d}{4} x^4 = \rho_i. 
\eeq

Rearranging the above equation in the form 
\beq
Ay^4 + By^3 + Cy^2 + Dy + E = 0, \nn
\eeq
we obtain the equation of curves, $y_k(x)$, $k = 1, 2, 3, 4$, for (anti-)thimbles as a function of $x$
\bseq
\label{eq:GA-thimble-dneq0}
\bea
y_{1, 2} &=& -\frac{B}{4A} - S \pm \frac{1}{2} \sqrt{-4S^2 - 2P + \frac{Q}{S}}, \\
y_{3, 4} &=& -\frac{B}{4A} + S \pm \frac{1}{2} \sqrt{-4S^2 - 2P - \frac{Q}{S}}, \\
P &=& \frac{8AC - 3 B^2}{8A^2}, \\
Q &=& \frac{B^3 - 4ABC + 8A^2D}{8A^3}, \\
R &=& \left( \frac{\Delta_1 + \sqrt{\Delta_1^2 - 4 \Delta_0^3}}{2}\right), \\
S &=& \frac{1}{2} \sqrt{-\frac{2}{3}P + \frac{1}{3a} \left( R + \frac{\Delta_0}{R} \right)}, \\
\Delta_0 &=& C^2 - 3BD + 12AE, \\
\Delta_1 &=& 2C^3 - 9BCD + 27B^2E + 27AD^2 - 72ACE.
\eea
\eseq

Although the solutions to the thimble equation given in Eq. \eqref{eq:GA-thimble-eq-dneq0} exist in the form of Eq. \eqref{eq:GA-thimble-dneq0}, there are a few caveats we would like to stress on. There are too many conditions to keep track of due to the requirement that $x, y \in \mathbb{R}$. (A visual summary of these conditions can be found in Ref. \cite{rees_graphical_1922}.) These conditions could potentially lead to discontinuities in the curve equations, $y_k(x)$, $k = 1, 2, 3, 4$, for the thimbles in Eq. \eqref{eq:GA-thimble-dneq0}. Further, the requirement of keeping track of these conditions manifests itself as the four solutions simultaneously being either the thimble or the anti-thimble depending on the region in the complex plane under consideration. We will refer to this as the `piecewise behavior' of the solutions since they appear themselves as piecewise thimbles/anti-thimbles/ghosts. Whether a solution shows piecewise behavior or not depends on the set of parameters $\{ a, b, c, d \}$.

\begin{figure*}[!htp]

\subfloat[Thimbles, anti-thimbles, and ghosts for all the critical points, $\phi_0$ and $\phi_\pm$, of the action.]{\includegraphics[width=5cm]{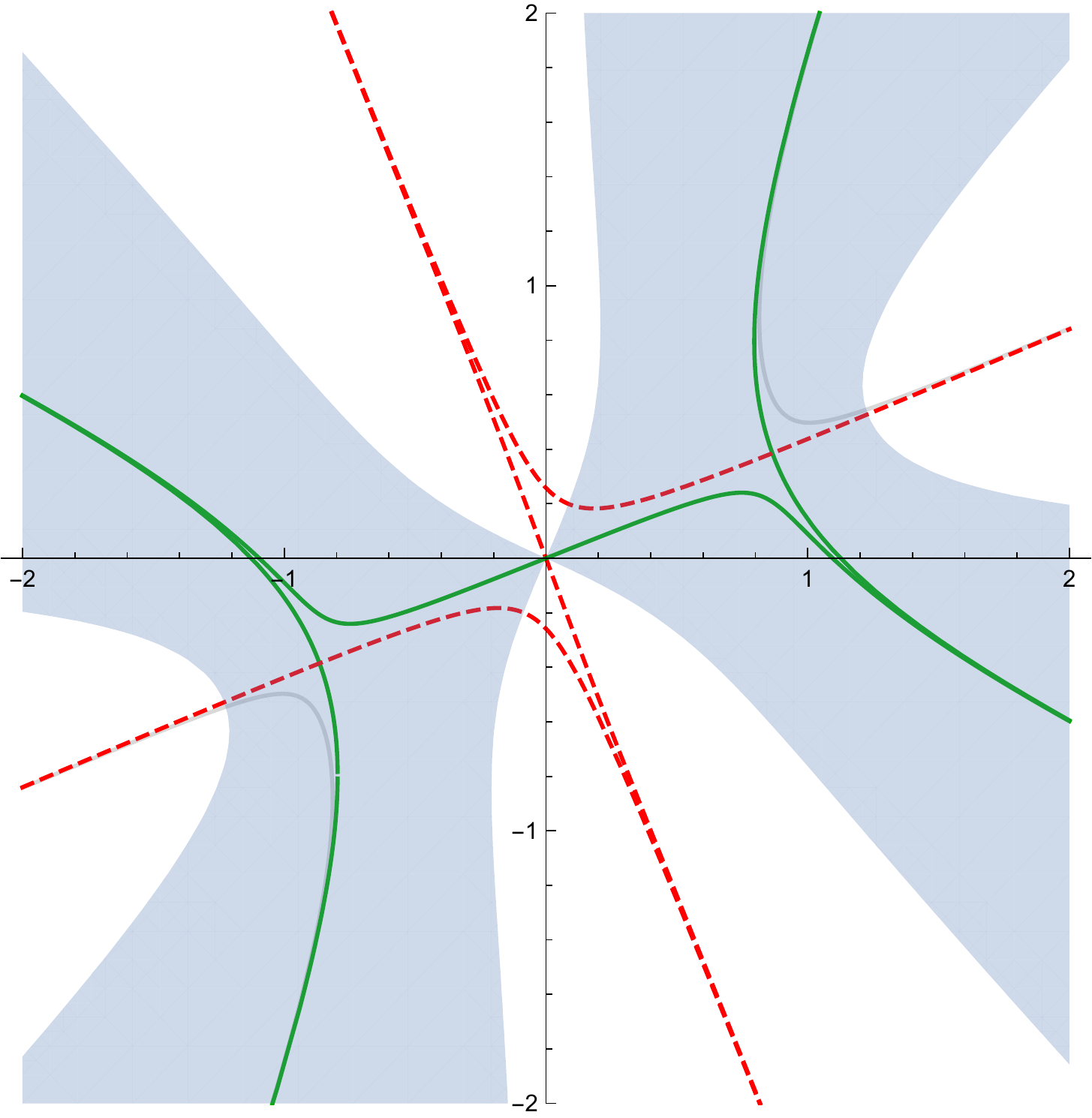}}\label{AllComplexFigure1}
$~~~$
\subfloat[Thimble, anti-thimble, and ghosts for the critical point $\phi_0$ of the action.]{\includegraphics[width=5cm]{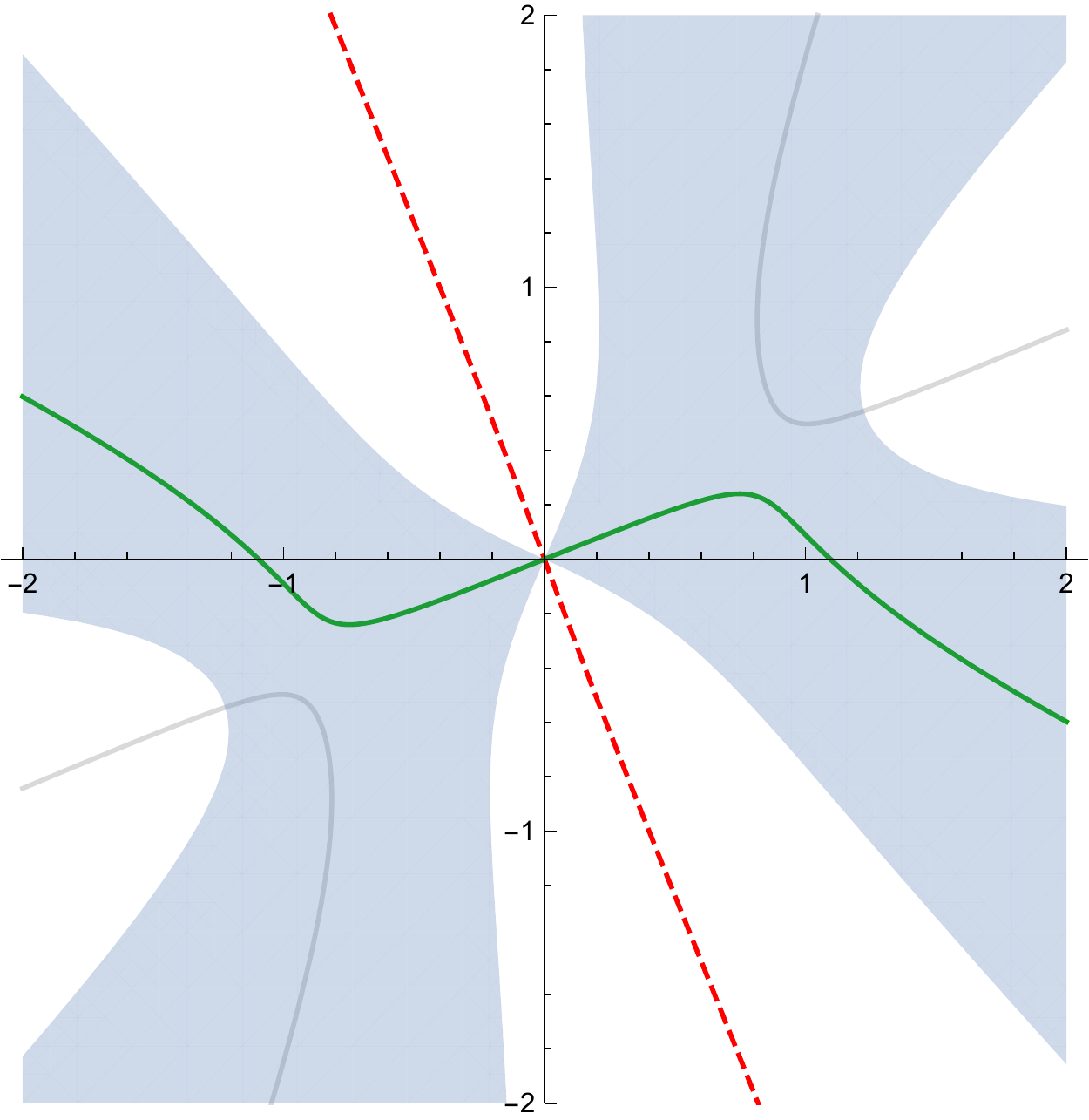}}\label{AllComplexFigure2}
$~~~$
\subfloat[Thimbles and anti-thimbles for the critical points $\phi_\pm$ of the action.]{\includegraphics[width=5cm]{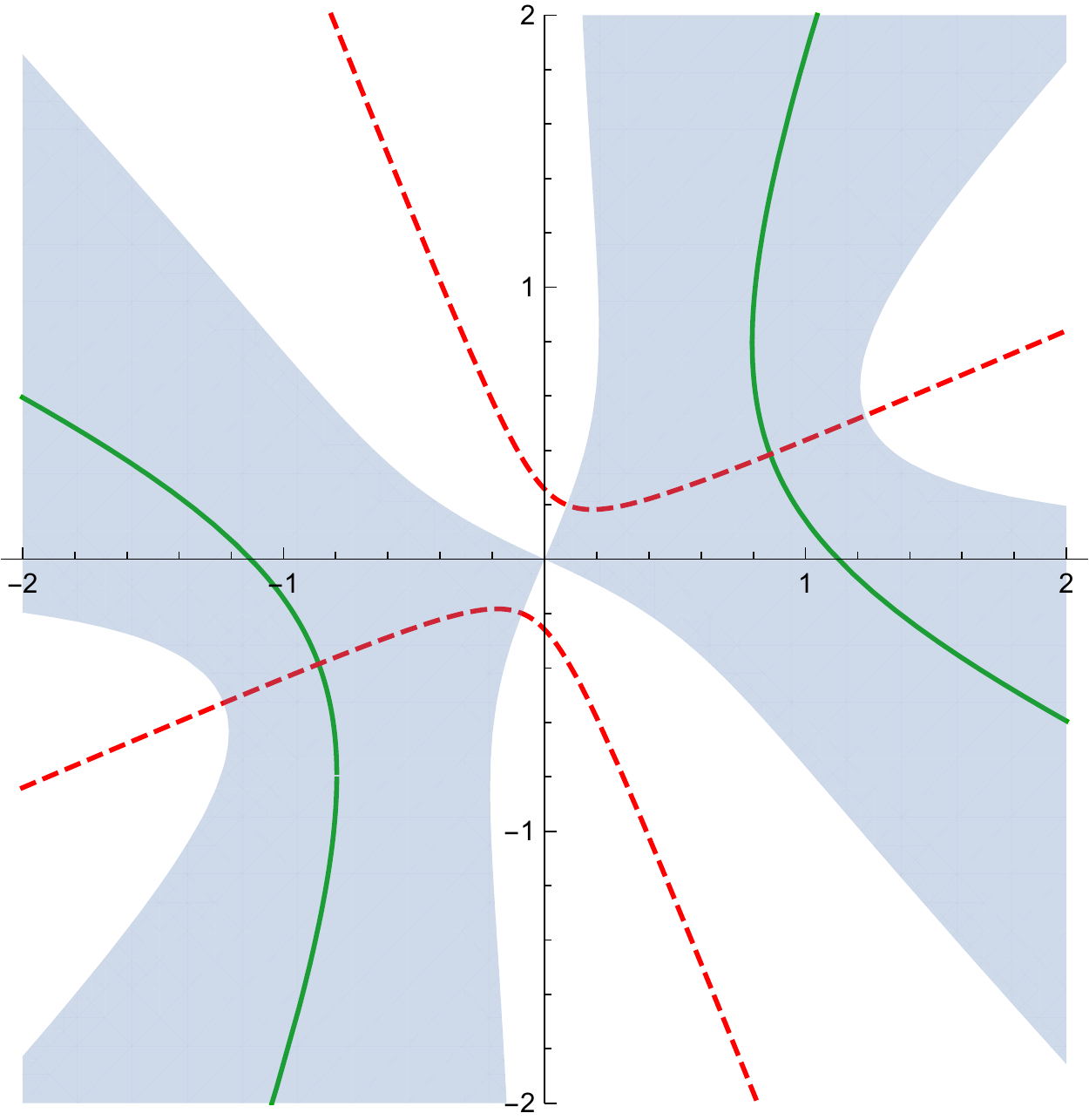}}\label{AllComplexFigure3}

\caption{(Color online). The solutions to the thimble equation given in Eq. \eqref{eq:thimble-equation}, corresponding to the critical points $\phi_0$ and $\phi_\pm$, for the parameters $\{a = 1$, $b = -0.9$, $c = 0$, $d = 1.5$, $h = 0\}$. In all the three figures, the green solid curves represent the thimbles, red dashed curves represent the anti-thimbles, and the grey solid curves represent the ghosts. The shaded regions represent the regions where $\text{Re}(S)\geq0$.}
\label{fig:GA-thimbles-dneq0-1}
\end{figure*}

\begin{figure*}[!htp]

\subfloat[Thimbles, anti-thimbles, and ghosts for all the critical points, $\phi_0$ and $\phi_\pm$, of the action.]{\includegraphics[width=5cm]{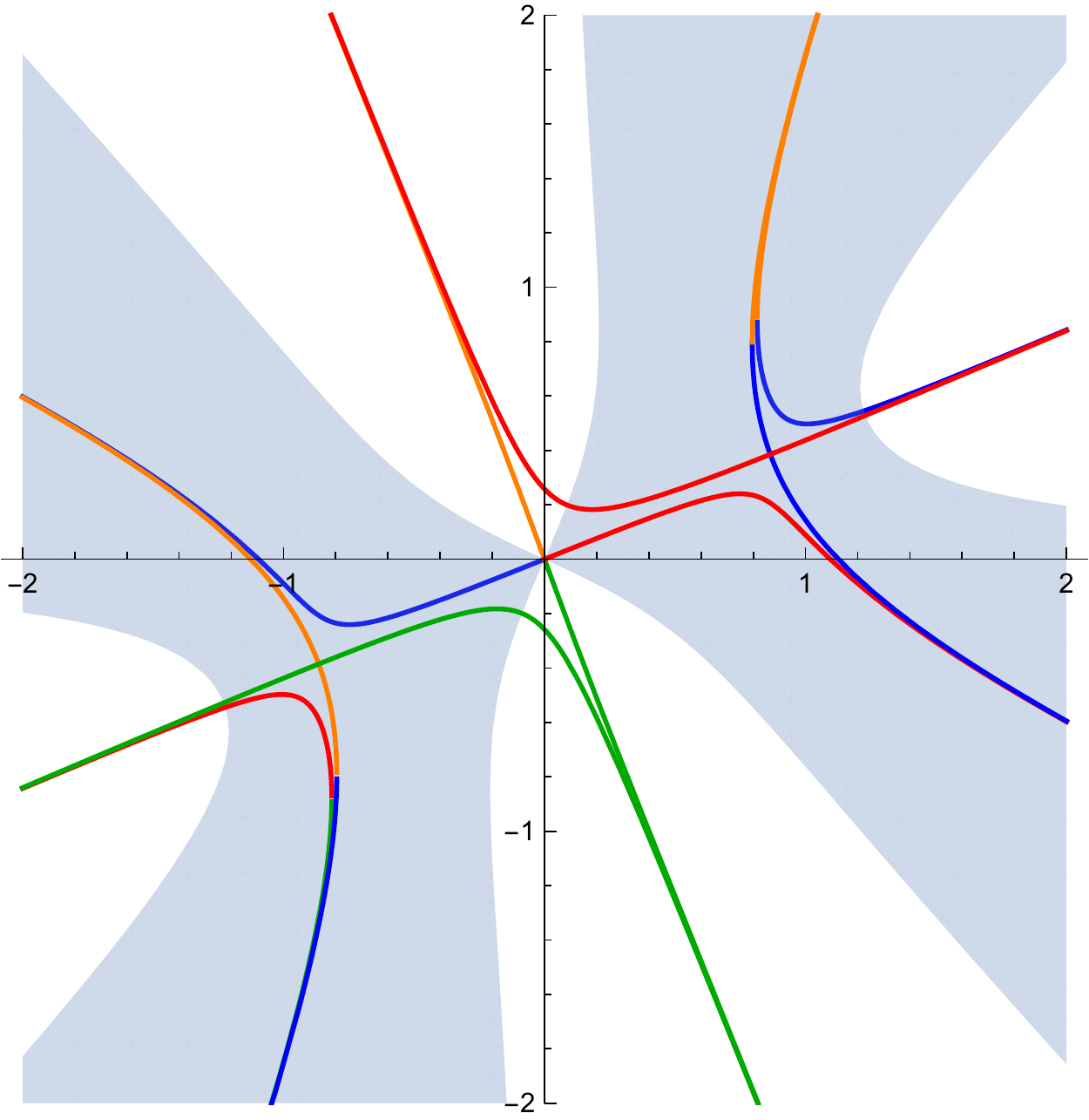}}\label{AllComplexFigure4}
$~~~$
\subfloat[Thimble, anti-thimble, and ghosts for the critical point $\phi_0$ of the action.]{\includegraphics[width=5cm]{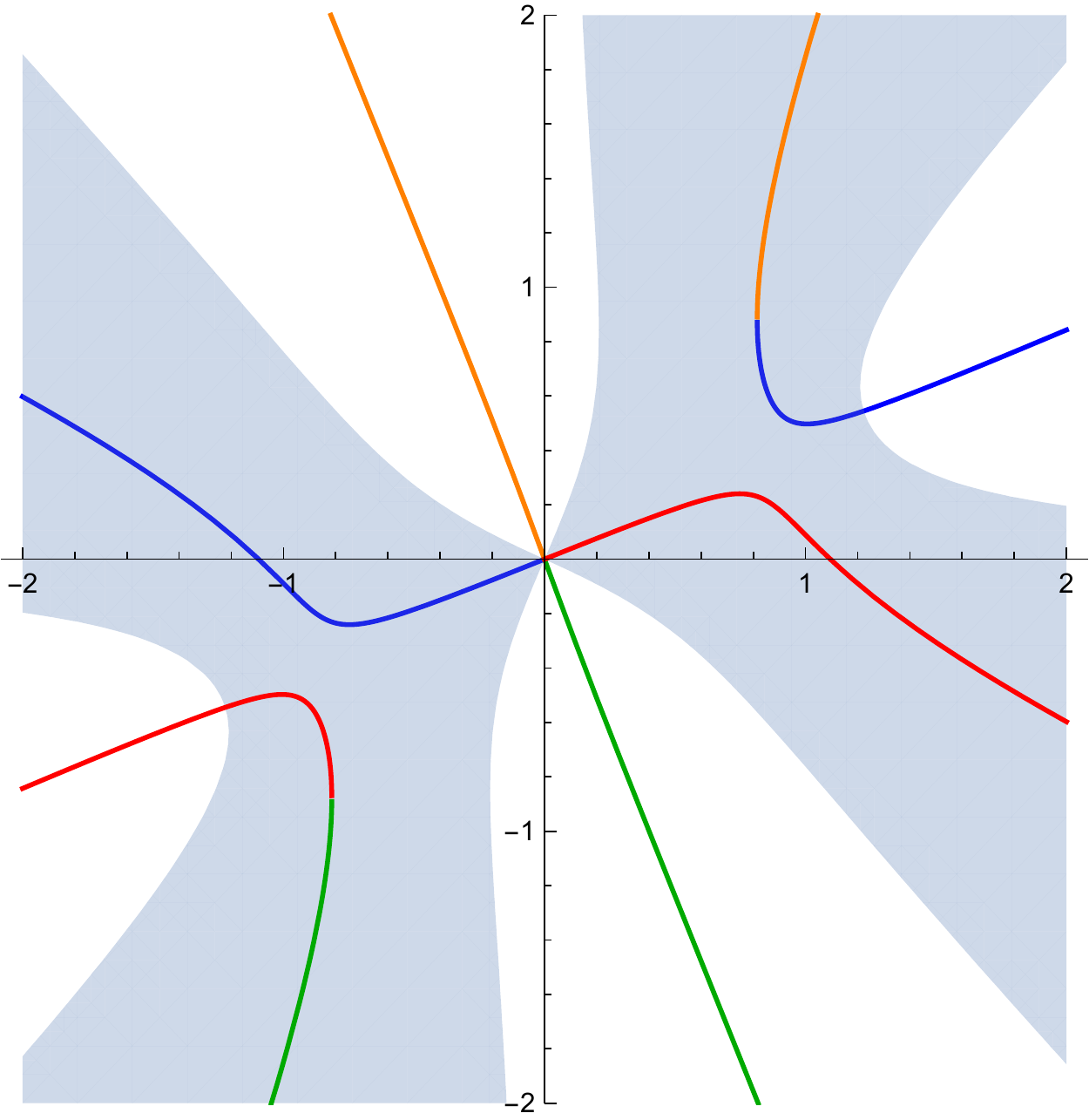}}\label{AllComplexFigure5}
$~~~$
\subfloat[Thimbles and anti-thimbles for the critical points $\phi_\pm$ of the action.]{\includegraphics[width=5cm]{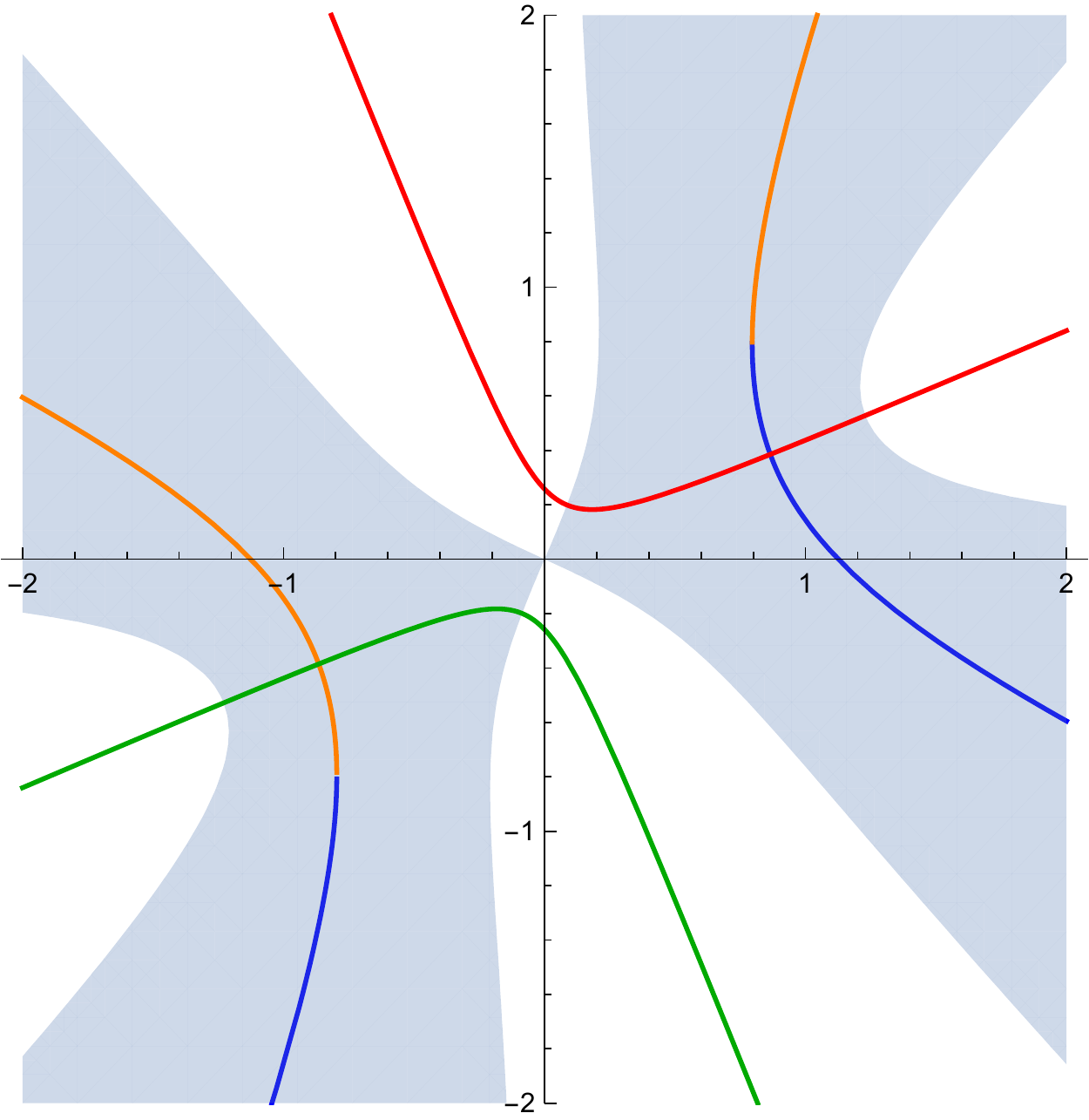}}\label{AllComplexFigure6}

\caption{(Colour online). Demonstration of the piecewise behavior by the solutions to the thimble equation given in Eq. \eqref{eq:thimble-equation}, corresponding to the critical points $\phi_0$ and $\phi_\pm$ of the action, for the parameters $\{a = 1$, $b = -0.9$, $c = 0$, $d = 1.5$, $h = 0\}$. In all the three figures, the blue solid curves correspond to solution $y_1$, orange solid curves to the solution $y_2$, green solid curves to the solution $y_3$ and the red solid curve to the solution $y_4$. The shaded regions denote the regions where $\text{Re}(S)\geq0$.}
\label{fig:GA-thimbles-dneq0-1-piecewise}
\end{figure*}

\begin{figure*}[htp]

\subfloat[Thimbles, anti-thimbles, and ghosts for all the critical points, $\phi_0$ and $\phi_\pm$, of the action.]{\includegraphics[width=5cm]{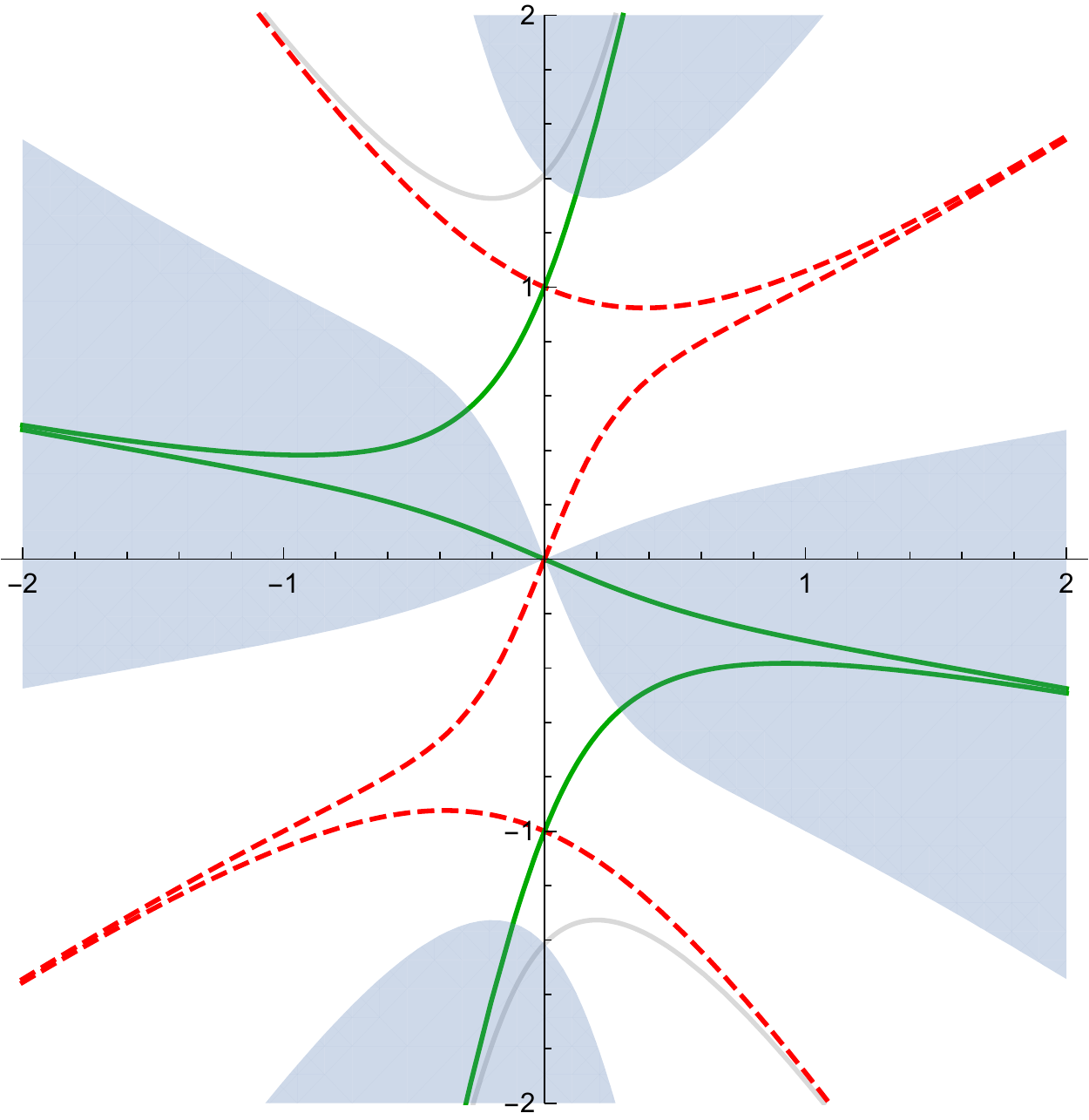}}\label{AllComplexFigure7}
$~~~$
\subfloat[Thimble, anti-thimble, and ghosts for the critical point $\phi_0$ of the action.]{\includegraphics[width=5cm]{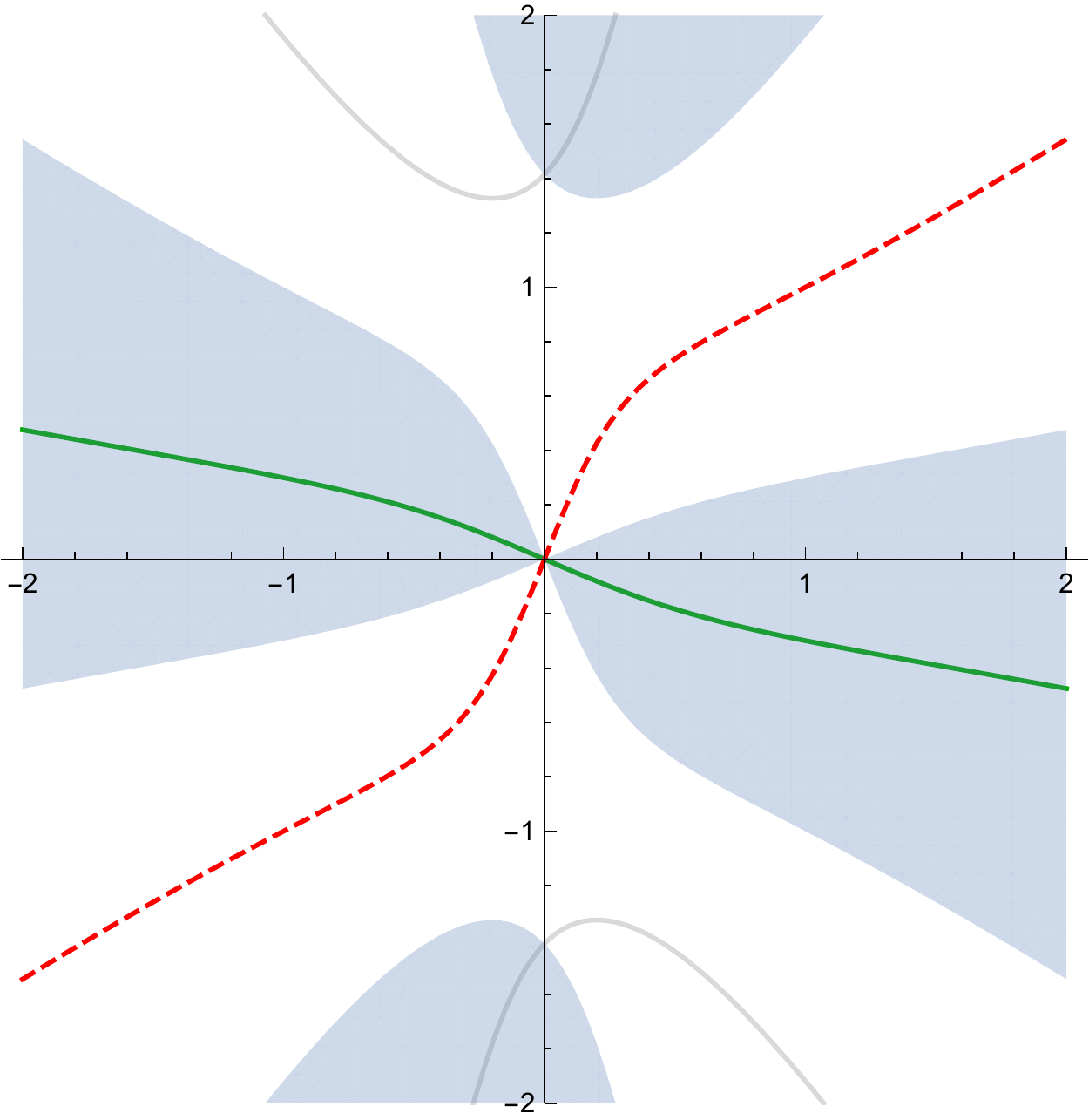}}\label{AllComplexFigure8}
$~~~$
\subfloat[Thimbles and anti-thimbles for the critical points $\phi_\pm$ of the action.]{\includegraphics[width=5cm]{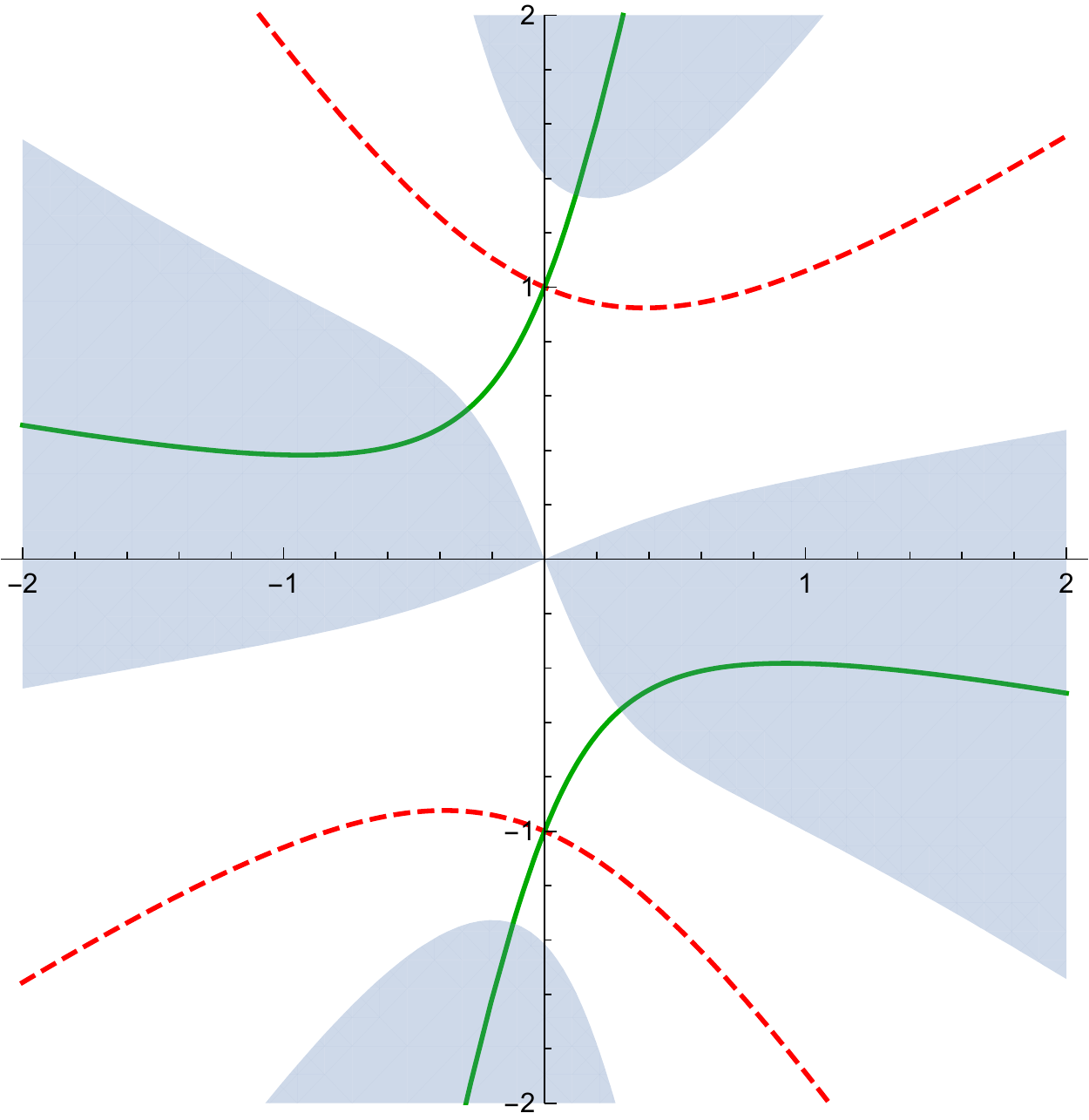}}\label{AllComplexFigure9}

\caption{(Colour online). The solutions to the thimble equation, Eq. \eqref{eq:GA-thimble-eq-dneq0}, corresponding to $\phi_0$ (Middle) and $\phi_\pm$ (Right), for the parameters $\{a = 1$, $b = 1$, $c = 1$, $d = 1$, $h = 0\}$. In all the three figures, the green solid curves represent the thimbles, red dashed curves represent the anti-thimbles, and the grey solid curves represent the ghosts. The shaded regions represent the regions where $\text{Re}(S)\geq0$.}
\label{fig:GA-thimbles-dneq0-2}

\end{figure*}

\begin{figure*}[!htp]

\subfloat[Thimbles, anti-thimbles, and ghosts for all the critical points, $\phi_0$ and $\phi_\pm$, of the action.]{\includegraphics[width=5cm]{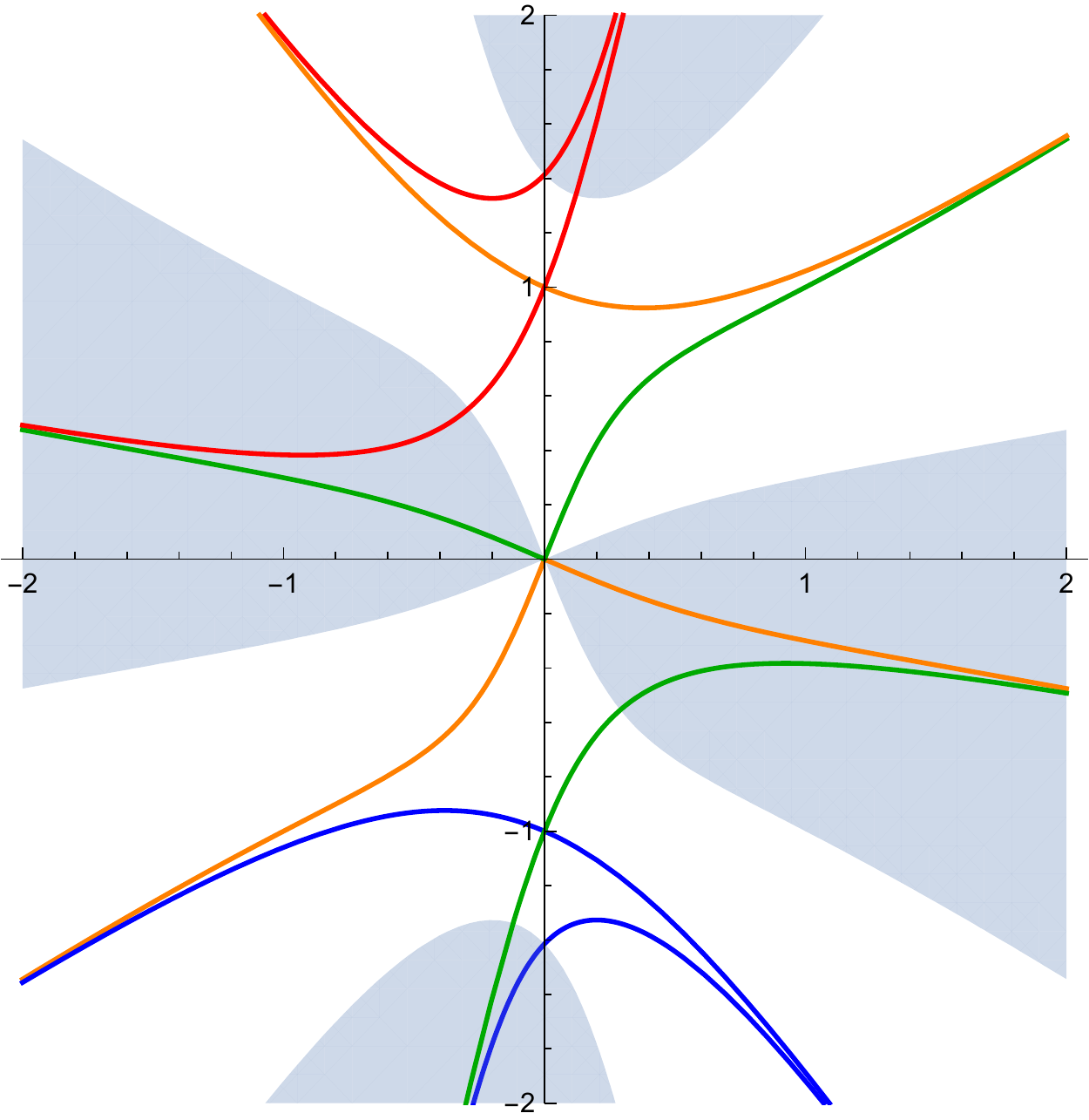}}\label{AllComplexFigure10}
$~~~$
\subfloat[Thimble, anti-thimble, and ghosts for the critical point $\phi_0$ of the action.]{\includegraphics[width=5cm]{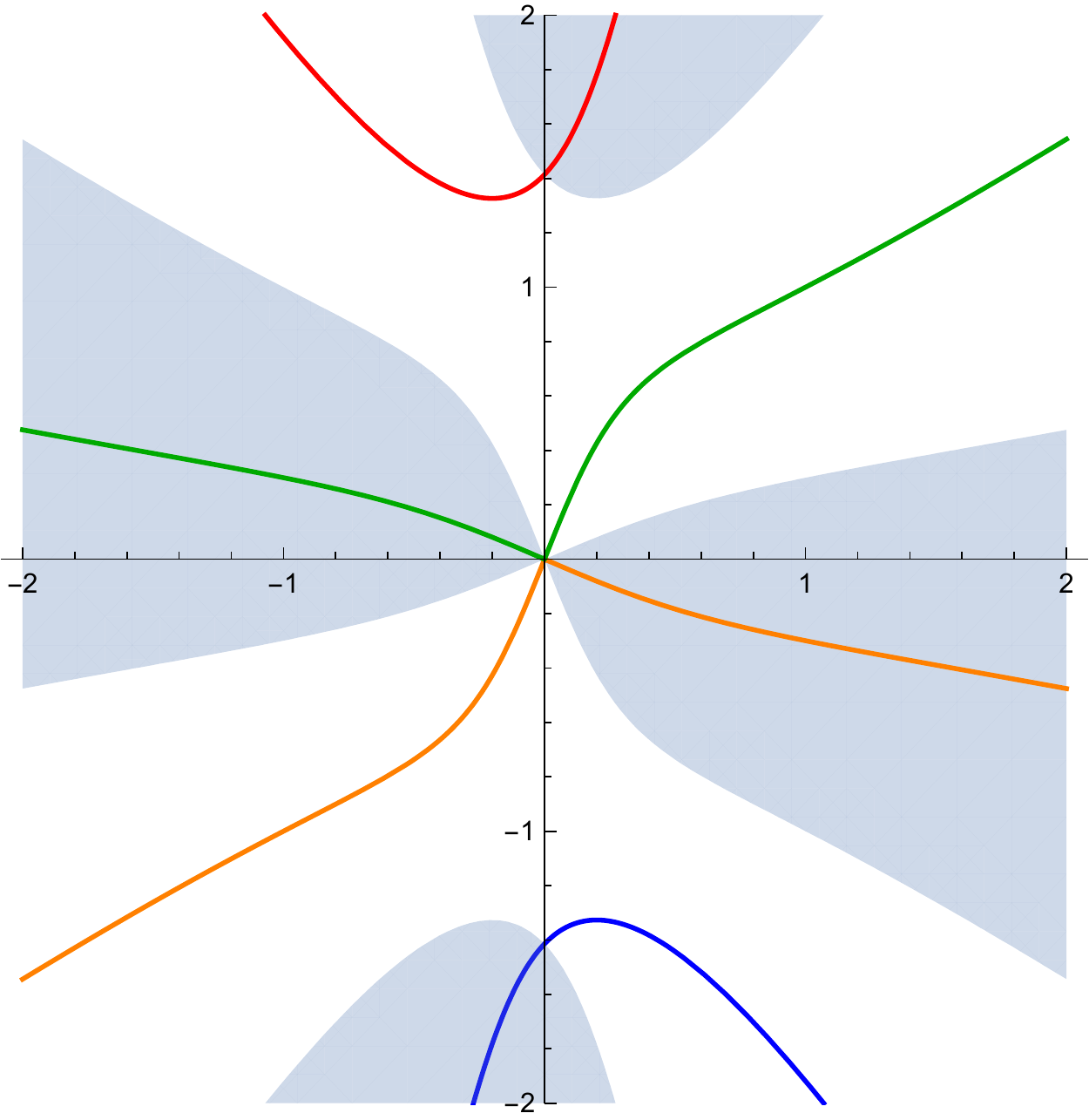}}\label{AllComplexFigure11}
$~~~$
\subfloat[Thimbles and anti-thimbles for the critical points $\phi_\pm$ of the action.]{\includegraphics[width=5cm]{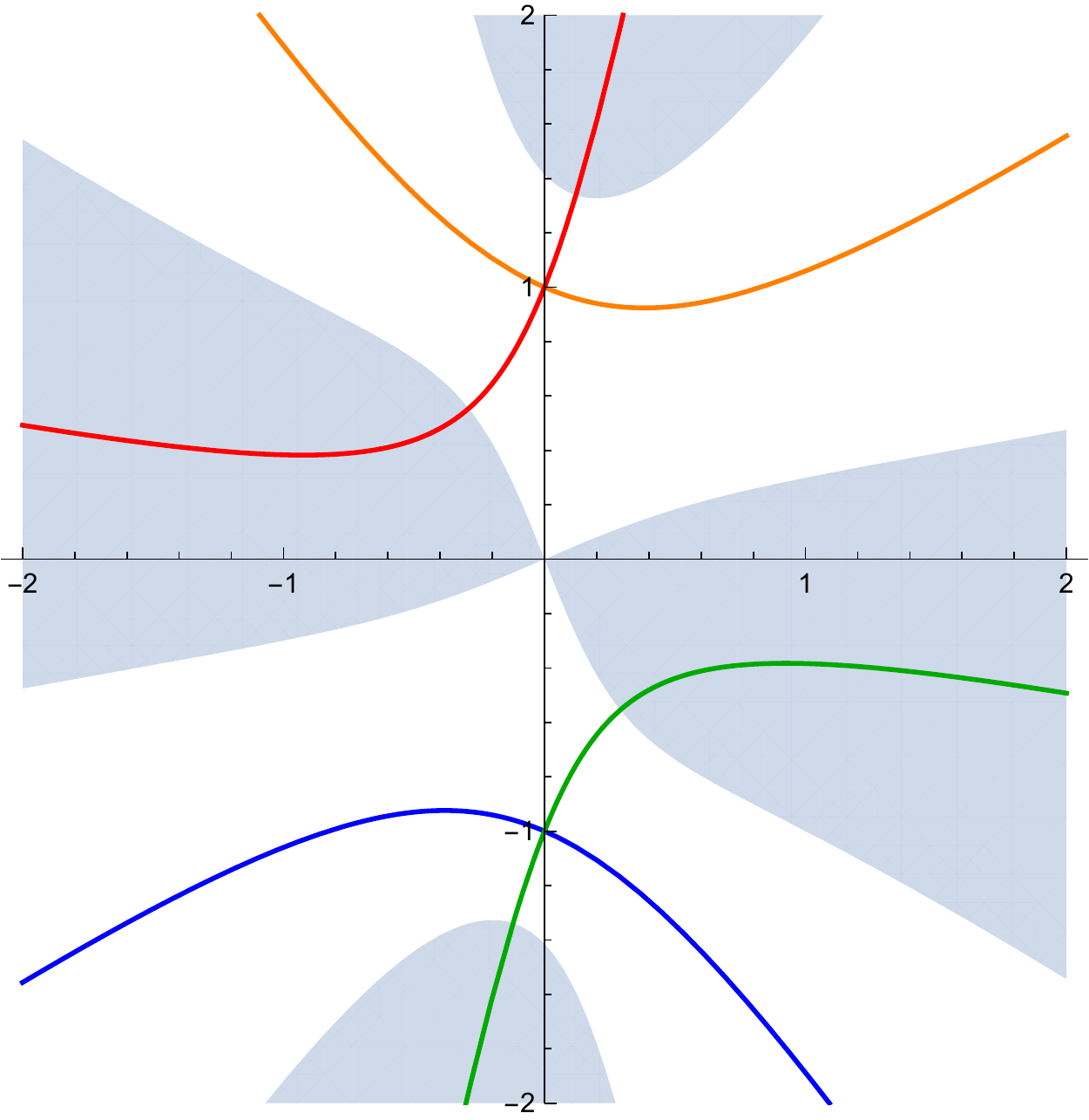}}\label{AllComplexFigure12}

\caption{(Colour online). Demonstration of piecewise behavior by the thimble solutions to Eq. \eqref{eq:thimble-equation}, corresponding to $\phi_0$ and $\phi_\pm$, for the parameters $\{a = 1$, $b = 1$, $c = 1$, $d = 1$, $h = 0\}$. In all the three figures, the blue solid curves correspond to solution $y_1$, orange solid curves to solution $y_2$, green solid curves to solution $y_3$, and red solid curves to solution $y_4$. The shaded regions denote the regions where $\text{Re}(S)\geq0$.}
\label{fig:GA-thimbles-dneq0-2-piecewise}
\end{figure*}

Let us consider the examples illustrated in Figs. \ref{fig:GA-thimbles-dneq0-1} and \ref{fig:GA-thimbles-dneq0-1-piecewise}. We see that the solution $y_1$ for $\phi_0$ gives the thimble for $x < 0$ and a ghost for $x > 0$, $y_2$ gives the anti-thimble for $x < 0$ and a ghost for $x > 0$, $y_3$ gives the anti-thimble for $x > 0$ and a ghost for $x < 0$, and $y_4$ for gives the anti-thimble for $x > 0$ and a ghost for $x < 0$. Similarly, for $\phi_\pm$, the solutions $y_1$ and $y_2$ give the thimble for both $x < 0$ and $x > 0$, and $y_3$ and $y_4$ give the thimble for both $x < 0$ and $x > 0$. However, they still exhibit the piecewise behavior. There definitely are parameter sets $\{ a, b, c, d\}$ for which the piecewise behavior might not be exhibited. One such case is when $\{a = 1$, $b = 1$, $c = 1$, $d = 1\}$; six of the eight solutions do not exhibit this behavior. (We show this in Figs. \ref{fig:GA-thimbles-dneq0-2} and \ref{fig:GA-thimbles-dneq0-2-piecewise}.) These cases, however, seem to be exceptions rather than the norm.

From the thimble/anti-thimble/ghost solutions given in Eq. \eqref{eq:GA-thimble-dneq0}, we see that obtaining the curves for the case $h \neq 0$ is straightforward. If $h$ is real, then $C$ changes from $(ax + cx^3)$ to $(h + ax + cx^3)$ while $E$ remains the same, except for the change in $\rho_i$. If $h$ is purely imaginary, then $C$ remains unchanged and $E$ gains an additional $hx$ term apart from the change to $\rho_i$. This situation also suffers from the issues discussed earlier for the case where $h$ was taken to zero while $d$ was non-zero. 

\subsection{Partition Function and Observables}
\label{ssec:Partition_Function_and_Observables}

Let us consider the action given in Eq. \eqref{eq:action} for the case $h = 0$. We have
\beq
S[\phi] = \frac{\sigma}{2} \phi^2 + \frac{\lambda}{4} \phi^4.
\eeq
This model is referred to as the quartic model and it represents the simplest nontrivial quantum field theory action. This model was studied extensively in the context of complex Langevin dynamics in Refs. \cite{Duncan:2012tc, Aarts:2013fpa, Aarts:2013uza}.

The $n$-point functions of the model can be constructed in the following way
\beq
\langle \phi^n \rangle = \frac{1}{Z} \int D\phi \; \phi^n e^{-S[\phi]}, ~~ Z = \int D\phi \; e^{-S[\phi]},
\eeq
with $Z$ denoting the partition function.

Now consider the following integral associated with the above action along the original integration cycle $\mathbb{R}$
\beq\label{eq:In}
I_n = \int_{-\infty}^{\infty} dx \; x^n \exp \left[ - \left( \frac{\sigma}{2} x^2 + \frac{\lambda}{4}x^4 \right) \right].
\eeq
Since the integrand is of odd parity under the exchange $x \rightarrow -x$ when $n$ is odd, the above integral is non-zero only for even values of $n$. We recover the partition function when $n = 0$, and the observables for the system are related to the above integral as 
\beq
\langle \phi^n \rangle = \frac{1}{Z} I_n.
\eeq

The exact result of the integral is known in terms of modified Bessel functions for the cases $n = 0$ and $n = 2$, for $\text{Re}(\sigma) > 0$ and $\text{Re}(\lambda) > 0$, as \cite{Abe:2016hpd} 
\bea
Z &=& \sqrt{ \frac{\sigma}{2 \lambda} }~e^{ \sigma^2/(8 \lambda) }~K_{1/4} \left( \frac{\sigma^2}{8 \lambda} \right), \\
\langle \phi^2 \rangle &=& \frac{\sigma}{4 \lambda} \frac{{\rm K}_{-3/4} \left( \frac{\sigma^2}{8 \lambda} \right) + {\rm K}_{5/4} \left( \frac{\sigma^2}{8 \lambda} \right)}{{\rm K}_{1/4} \left( \frac{\sigma^2}{8\lambda}  \right)} - \frac{\sigma}{2\lambda} - \frac{1}{\sigma}.
\eea
Here $K$ is the modified Bessel function of the second kind. In the case where ${\rm Re}(\sigma) < 0$, we replace $K$ in $Z$ with $I$, the modified Bessel function of the first kind.

Integrating Eq. \eqref{eq:In} by parts, rearranging, and dividing by $Z$, we obtain a recursion relation for observables of the theory
\beq
\label{eq:recursion}
(2n+1) \langle \phi^{2n} \rangle - \sigma \langle \phi^{2(n+1)} \rangle - \lambda \langle \phi^{2(n+2)} \rangle = 0.
\eeq

Thus, since the closed-form expressions for the partition function and the observable $\langle \phi^2\rangle$ are known, all observables of the theory are known and can be written in terms of the two using Eq. \eqref{eq:recursion}. The relation could potentially be used to determine the partition function of the action with sources. For nonzero $h$ the partition function is given by
\beq
Z_{\rm sources} = \int_{-\infty}^{\infty} dx \; e^{-hx}~\exp \left[ - \left( \frac{\sigma}{2}x^2 + \frac{\lambda}{4} x^4 \right) \right].
\eeq

Taylor expanding the first exponential, we get
\beq
\label{eq:partition-function-sources}
Z_{{\rm sources}} = \sum_{n=0}^\infty \frac{h^{2n}}{(2n)!}I_{2n},
\eeq
and from the recurrence relation derived above, $Z_{\rm sources}$ can be written solely in terms of $I_0$ and $I_2$. 
\section{Determining the Intersection Numbers}
\label{sec:Phase_Transition_Boundaries}

The intersection number $n_i$, defined in Eqs. \eqref{eq:z-integral} and \eqref{eq:int-number-delta}, being an integer, greatly controls the behavior of the partition function and observables of the model. 

As the parameters of the action are changed, the intersection number corresponding to a critical point could potentially change, which results in an abrupt change in the values of the partition function and the observables of the system. The most dramatic among these is the case when the intersection number takes the value zero, and this in turn results in the corresponding critical point not contributing to the dynamics of the system. This change in the intersection number is referred by the name {\it Stokes phenomena}\footnote{An alternative, and equivalent, definition used frequently in the literature in the context of integration by the method of steepest descent is the change in the asymptotic formula for the same analytic function when the parameters of the function are changed \cite{Aniceto:2018bis, riley_hobson_bence_2002}.} \cite{Witten:2010cx, Fukushima:2015qza}. This phenomenon points at the existence of quantum phase transitions in the system.

We note that the abrupt changes in the values of the partition function and observables occur as a function of one or more non-thermal parameters in the model. These type of transitions at zero temperature are called quantum phase transitions or quantum critical points. We can capture quantum phase transitions in the system by observing the appearance of non-analyticity in the observables as a function of one or more non-thermal control parameters. In our case, the control parameters belong to the set $\{ \sigma, \lambda, h \}$. For more details on quantum phase transitions see Ref. \cite{sachdev2011quantum}.

The power of using Eq. \eqref{eq:thimble-equation} to solve for (anti-)thimbles is the fact that it captures the information about these intersection numbers. Using this equation, we can look for the values of the control parameters around which the intersection number jumps. This, in turn, allows us to predict the boundaries of these phase transitions. One of our main results will be the analytic expressions for the combined intersection number of thimbles and anti-thimbles of the simplest nontrivial quantum field theory: a scalar-field theory in zero-spacetime dimensions containing a quartic interaction term and a source term.

When the parameter $c$ is positive, to arrive at these expressions, we use the fact that the original integration cycle $\mathbb{R}$ corresponds to $y = 0$. Substituting this in Eq. \eqref{eq:thimble-equation}, we obtain a polynomial equation in $x$ of degree four or lower. Looking at the number of real solutions to the polynomial equation (remember, $x \in \mathbb{R}$) gives us the information about the number of times the thimbles and anti-thimbles\footnote{The number of solutions could potentially also contain information about the number of times a ghost solution intersects the original integration cycle. However, we have not come across a situation where a ghost solution intersects the real line. This is explained by the observation that a ghost solution always has one end inside the region of stability and the other end inside the region of instability. This, along with the fact that these curves do not intersect either the thimble or the anti-thimble of the same critical point tells us that the ghosts are always away from the real line.} intersect the original integration cycle. When $c$ is negative, (this is the case for the action with ${\cal PT}$ symmetry) we can substitute Eq. \eqref{eq:pt-symmetry-cycle} in Eq. \eqref{eq:thimble-equation} and repeat the above analysis. 

\subsection{A Simple Demonstration using Airy Integral}

Before we present our results for the action given in Eq. \eqref{eq:action}, as a primer, let us look at quantum phase transitions in a model containing the Airy integral as the action. 

Consider the following integral
\beq
\label{eq:airy-integral}
\text{Ai}(u) = \int_{-\infty}^{\infty} \text{exp}\left(i \left\{ \frac{x^{3}}{3} + u x\right\} \right),
\eeq
where we restrict $u$ to take real values. This integral is equivalent to taking our action (after continuation to the complex plane) as
\beq
S[\phi] = -i \left\{ \frac{\phi^{3}}{3} + u \phi \right\}.
\eeq

There are two critical points of this action, namely $\phi_{\pm} = \pm i \sqrt{u}$. At these critical points, the action takes the values
\beq
S[\phi_{\pm}] = \pm \frac{2}{3} u^{3/2}. 
\eeq
The imaginary part of the action is
\beq
\label{eq:airy-imag}
\text{Im}(S[\phi]) = \frac{3xy^{2}-x^{3}}{3} - u x.
\eeq

To look for phase transition boundaries, we look for the number of real solutions to the equation
\beq
\label{eq:airy-thimble}
\text{Im}(S[\phi]) \Big| _{\text{along } \mathbb{R}} = \text{Im}S[\phi_{\pm}],
\eeq
which is equivalent to putting $y=0$ in Eq. \eqref{eq:thimble-equation}. Thus we look for real solutions to the equation
\beq
\frac{-x^{3}}{3} - u x = 0.
\eeq

For cubic equations, the number of solutions depends only on the sign of the discriminant, which for the above equation is
\beq
\Delta = -\frac{4}{3} u.
\eeq
When the discriminant is negative, the number of real solutions to the cubic equation is one, and when the discriminant is positive, the number of solutions is three. Thus we expect a phase transition at $u = 0$. In fact, this phase transition coincides with the change in the asymptotic expansion of Ai$(u)$.

There is a very subtle detail that must be noted. For real $u$, it is not possible to find the thimble for $\phi_{-}$ and the anti-thimble for $\phi_{+}$ without deforming $u$ into the complex plane as $u + i \epsilon$ for small $\epsilon$. This occurs because when $u$ is real, the two critical points are always connected by a {\it Stokes ray} \cite{Witten:2010cx}. More specific to our method, this problem arises because when $u$ is real, Eq. \eqref{eq:airy-thimble} has a vanishing right hand side. This leads to four curves being described by a polynomial equation of degree three, which cannot occur unless the critical points are connected by a flow. This is referred to as being connected by a Stokes ray. The deformation $u \rightarrow u + i \epsilon$ moves the critical points away from the Stokes ray, allowing us to find the thimbles and anti-thimbles.

If we take $u$ in Eq. \eqref{eq:airy-integral} to be complex we will arrive at a slightly more complicated situation. We will get a similar phase transition structure, where the phase boundary is $|\text{arg}(x)| = 2\pi/3$. This was shown explicitly using the Lefschetz thimble formalism by Tanizaki in Ref. \cite{Tanizaki:2015gpl}.

We now move on to our action with quartic interactions. Due to the differences in algebraic calculations and physical interpretations, we divide our results into multiple cases, and provide the detailed calculations that led to the results in Appendix \ref{sec:Expressions_for_Boundaries_of_Phase_Transitions}. 

\subsection{Quartic Model Without Source Term} 

\subsubsection{Real Coupling}

Let us consider the case $h = 0$, $\sigma, \lambda \in \mathbb{C}$, ${\rm Re}(\lambda) \geq 0$, and $d = 0$. The thimble equation, Eq. \eqref{eq:thimble-equation}, gives a quadratic in $x$, from which the intersection numbers can trivially be found based on the conditions given below

\beq
\label{eq:ni-d=0}
n_i \left\{
\begin{array}{ll}
     = 1 & {\rm ~~~~when~} i = 0, \forall ~~ \sigma, \lambda, \\
     \leq 1 & {\rm ~~~~when~} i = \pm, \frac{a}{c} < 0, \\
     = 0 & {\rm ~~~~when~} i = \pm, \frac{a}{c} > 0. \\
\end{array} 
\right. 
\eeq

This is the easiest of the cases that have been considered. In further analyses, the possibility of $d = 0$, where the equations reduce to a quadratic instead of the original quartic in $x$, is not considered since repeating the calculation by requiring $d = 0$ is straightforward.

\subsubsection{Complex Coupling} 

Upon relaxing the condition on $d$ while maintaining $h = 0$, $\sigma, \lambda \in \mathbb{C}$, and ${\rm Re}(\lambda) \geq 0$, the polynomial obtained from Eq. \eqref{eq:thimble-equation} is a bi-quadratic in $x$. 

Let us define the variables $\Delta$, $\Pi$ and $\Sigma$, which are related to the discriminant, product of roots, and sum of roots, respectively as outlined in Appendix \ref{sec:Expressions_for_Boundaries_of_Phase_Transitions}, as 
\bseq
\label{eq:exp-for-cond-h=0-dneq0}
\bea
\Delta &=& \frac{(bc-ad)^2}{(c^2+d^2)}, \\
\Pi &=& \frac{d(b^2-a^2)+2abc}{d(c^2+d^2)}, \\
\Sigma &=& \frac{b}{d}.
\eea
\eseq

Then the intersection number for the critical point $\phi_0$ is determined using the conditions in Table \ref{tab:hzero-dnotzero-phi0}, and the intersection number for the critical points $\phi_\pm$ is determined using the conditions in Table \ref{tab:hzero-dnotzero-phipm}. 

\begin{table}[!htp]
\bec
\begin{tabular}{ c  c }
\hline
\hline
Condition & $~~$Intersection \\
 & number\\
\hline
$\Sigma < 0$ & $\leq 3$ \\
$\Sigma \geq 0$ & $=1$ \\
\hline
\hline
\end{tabular}
\caption{Constraints on the intersection number for the critical point $\phi_0$ when $h = 0$, $\sigma, \lambda \in \mathbb{C}$, ${\rm Re}(\lambda) \geq 0$, and $d \neq 0$.}
\label{tab:hzero-dnotzero-phi0}
\eec
\end{table}

There are two comments to be made about these results. First, in both the Tables \ref{tab:hzero-dnotzero-phi0} and \ref{tab:hzero-dnotzero-phipm} (and later), we have extensively used the fact that (anti-)thimbles pass through the corresponding critical points. Further, in the situations discussed in this section, the (anti-)thimbles are not connected by the same flow equation, except for points in the parameter space at which the intersection number changes. Thus we have also used the fact that a (anti-)thimble of a particular critical point does not pass through any other critical point. Second, if a condition given in these tables does not provide any condition for a specific relation between the parameters (for instance, $\Pi$ and $\Delta$ in Table \ref{tab:hzero-dnotzero-phi0}), it is to be understood that the value of that particular relation does not affect the intersection number.

\bec
\begin{table}[!htp] 
\begin{tabular}{ c  c }
\hline
\hline
Condition & Intersection \\
 & number \\
 \hline
 $\Delta > 0$, $\Pi > 0$, $ \Sigma < 0$ & $\leq4$ \\
 
 $\Delta > 0$, $\Pi = 0$, $\Sigma < 0$ & $\leq2$ \\
 $\Delta > 0$, $\Pi < 0$ & $\leq2$ \\
 $\Delta = 0$, $\Sigma < 0$ & $\leq2$ \\
 
 $\Delta > 0$, $\Pi > 0$, $\Sigma \geq 0$ & $=0$ \\
 $\Delta > 0$, $\Pi = 0$, $\Sigma \geq 0$ & $=0$ \\
 $\Delta = 0$, $\Sigma \geq 0$ & $=0$ \\
\hline
\hline
\end{tabular}
\caption{Constraints on the intersection number for the critical points $\phi_\pm$ when $h = 0$, $\sigma, \lambda \in \mathbb{C}$, ${\rm Re}(\lambda) \geq 0$, and $d \neq 0$.}
\label{tab:hzero-dnotzero-phipm}
\end{table}
\eec

As an illustration, let us determine the boundary at which the Stokes phenomena occurs for the choice of constants $a = 1$, $c = 0$, and $d = 1.5$, as derived by Fukushima and Tanizaki in Ref. \cite{Fukushima:2015qza}\footnote{Note that the convention for constants used in Ref. \cite{Fukushima:2015qza} is slightly different but nonetheless, the results remain the same.}. Conditions for $\Pi$ given in Table \ref{tab:hzero-dnotzero-phipm} imply that all three thimbles to contribute when $b \in (-\infty, -1) \cup (1, \infty)$. Conditions for $\Sigma$ in Tables \ref{tab:hzero-dnotzero-phi0} and \ref{tab:hzero-dnotzero-phipm} further require $b < 0$, which implies that when $b \in (-\infty,-1)$, all three thimbles contribute, and that Stokes phenomena is observed around the critical coupling $b = b_c = -1$.

\subsection{Quartic Model With Source Term} 

\subsubsection{Real Source Parameter}

We now relax the condition on $h$ to $h\in\mathbb{R}$. The obtained equation, like the previous case, is a bi-quadratic but with a change to the part independent of $x$. 

Again let us introduce the variables $\Delta$, $\Pi$ and $\Sigma$ as
\beq
\label{eq:hreal-conditions}
\Delta = b^2 + 4 d \rho_i, ~~ \Pi = \frac{\rho_i}{d}, ~~ \Sigma = \frac{b}{d}.
\eeq
Here $\rho_i$ is the imaginary part of the action, as defined in Eq. \eqref{eq:imag-part-action-i}. The intersection number for each critical point $\phi_i$ is now determined by the conditions given in Table \ref{tab:hreal}.

\bec
\begin{table}[!htp] 
\begin{tabular}{ c  c c c c}
\hline
\hline
Condition & Intersection \\
 & number \\
 \hline
 $\Delta >0$, $\Pi >0$, $\Sigma<0$ & $\leq4$ \\
 
 $\Delta >0$, $\Pi =0$, $\Sigma<0$ & $\leq3$ \\
 
 $\Delta >0$, $\Pi <0$ & $\leq2$ \\
 $\Delta =0$, $\Sigma<0$ & $\leq2$ \\
 
 $\Delta >0$, $\Pi =0$, $\Sigma>0$ & $\leq1$ \\
 $\Delta =0$, $\Sigma=0$ & $\leq1$ \\
 
 $\Delta >0$, $\Pi>0$, $\Sigma>0$ & $=0$ \\
 $\Delta =0$, $\Sigma>0$ & $=0$ \\
 $\Delta <0$ & $=0$ \\
\hline
\hline
\end{tabular}
\caption{Constraints on the intersection number for the critical points $\phi_0, \phi_\pm$ when $h \neq 0, h \in \mathbb{R}$, $\sigma, \lambda \in \mathbb{C}$, ${\rm Re}(\lambda) \geq 0$, and $d \neq 0$.}
\label{tab:hreal}
\end{table}
\eec

For the situation where $\Delta > 0$, $\Pi \geq 0$, $\Sigma = 0$, the intersection number depends on the critical point under question. For $\phi_0$, the intersection number will be equal to one, while for $\phi_\pm$, the intersection number is zero.

\subsubsection{Imaginary Source Parameter} 

Let us consider the case when the source parameter is purely imaginary. Defining
\bea
\Delta &=& -\frac{1}{16} \left( 64d^3\rho_i^3 + 32b^2d^2 \rho_i^2 + 72bd^2h^2 \rho_i + 27d^2h^4 + 4b^4d \rho_i + 2b^3dh^2 \right), \\
\Delta_0 &=& \frac{1}{4} \left( b^2 - 12d \rho_i \right), \\
P &=& bd, \\
Q &=& -\frac{1}{4} \left( 4d \rho_i + b^2 d^2 \right), \\
R &=& \frac{d^2h}{2},
\eea
we obtain the conditions on the intersection number. They are provided in Table \ref{tab:himag}.

\bec
\begin{table}[!htp] 
\begin{tabular}{ c  c }
\hline
\hline
Condition & Intersection \\
 & number \\
 \hline
 
 $\Delta>0$, $P<0$, $Q<0$ & $\leq4$\\
 
 $\Delta=0$, $P<0$, $Q<0$, $\Delta_0\neq0$ & $\leq3$ \\
 
$\Delta=0$, $\Delta_0=0$, $Q\neq0$ & $\leq2$ \\
$\Delta=0$, $Q=0$, $P<0$ & $\leq2$\\
 $\Delta<0$ & $\leq2$ \\
 
 $\Delta=0$, $Q>0$ & $\leq1$ \\
 $\Delta=0$, $P>0$ & $\leq1$ \\
 (given $Q\neq0$ or $R\neq0$) & \\
 $\Delta=0$, $\Delta_0=0$, $Q=0$ & $\leq1$ \\
 
 $\Delta>0$, $P>0$, $Q>0$ & $=0$ \\
 $\Delta=0$, $P>0$, $Q=0$, $R=0$ & $=0$\\
 \hline
\hline
\end{tabular}
\caption{Constraints on the intersection number for $\phi_0, \phi_\pm$ when $h \in \mathbb{C}, \text{Re}(h) = 0$, $\sigma, \lambda \in \mathbb{C}$, ${\rm Re}(\lambda) \geq 0$, and $d \neq 0$.}
\label{tab:himag}
\end{table}
\eec

\subsection{Theory With $\mathcal{PT}$ Symmetry}

Let us look at zero-dimensional actions that possess the so-called ${\cal PT}$-symmetry, where ${\cal P}$ is the parity symmetry and ${\cal T}$ is the time reversal invariance. The motivation for considering ${\cal PT}$-symmetric theories is the following. In Ref. \cite{PhysRevD.62.085001} it was shown that imposing ${\cal PT}$-symmetric boundary conditions on the functional-integral representation of the four-dimensional $- \lambda \phi^4$ theory gives a spectrum that is bounded below. Such an interaction leads to a quantum field theory that is perturbatively renormalizable and asymptotically free, with a real and bounded-below spectrum. These properties suggest that a $- \lambda \phi^4$ quantum field theory might be useful in describing the Higgs sector of the Standard Model. We hope that our investigations in zero dimensions would serve as a starting point for exploring the thimble structures of these type of theories in higher dimensions.

In zero-spacetime dimensions, any {\it real} function of $ix$ is symmetric under ${\cal PT}$ transformation \cite{Bender:2007nj}. That is, our action should be of the form\footnote{We have only considered polynomials in $ix$ but any function with real powers of $ix$ is ${\cal PT}$-symmetric.}
\beq
\label{eq:PT-symm}
S = \sum_n - \lambda_n (ix)^n,
\eeq
with $n$ denoting integers and $\lambda_n$ representing real numbers.

Comparing Eq. \eqref{eq:PT-symm} with Eq. \eqref{eq:action}, we see that Eq. \eqref{eq:PT-symm} corresponds to the case with $h \in \mathbb{C}, {\rm Re}(h) = 0$, and $\sigma = a, \lambda = c \in \mathbb{R}$, such that $c < 0$. This is equivalent to replacing $h \rightarrow i h$ and $c \rightarrow -c$ in Eqs. \eqref{eq:action}, \eqref{eq:critical-points} and \eqref{eq:imag-part-action-i}, and maintaining $h \in \mathbb{R}$, $c > 0$. This leads to the following critical points of the action
\bseq
\bea
\phi_0 &=& -\frac{ih}{a} + \mathcal{O}(h^3), \\ 
\phi_\pm &=& \pm \sqrt{\frac{a}{c}} - \frac{ih}{2a}\pm \frac{3h^2}{8}\sqrt{\frac{c}{a^5}} + \mathcal{O}(h^3).
\eea
\eseq

The values of $\rho_i$ depend on the value of $a$. We have
\beq
\rho_i = \left\{
\begin{array}{ll}
      0 & {\rm when~}  a \leq 0, ~\forall ~i, \\
      0 & {\rm when~}  a > 0, ~i = 0, \\
      + h \sqrt{\frac{a}{|c|}} & {\rm when~}  a > 0, ~i = +, \\
      - h \sqrt{\frac{a}{|c|}} & {\rm when~}  a > 0, ~i = -. \\
\end{array} 
\right. 
\eeq

Let us explore the case when $a$ is positive. In this situation we obtain a set of quadratic equations. Solving for each sector in Eq. \eqref{eq:pt-symmetry-cycle}, and combining the results, we obtain the conditions in Tables \ref{tab:pt-symmetry-phi0} and \ref{tab:pt-symmetry-phipm}, where we have defined
\bea
\Delta_R &=& h^2+4a\rho_i, \\
\Delta_L &=& h^2-4a\rho_i, \\
\Sigma &=& \frac{h}{a}, \\
\Pi &=& \frac{\rho}{a}.
\eea

Combining the intersection numbers for each sector is highly non-trivial, and more information on how they were combined can be found in Appendix \ref{sec:Expressions_for_Boundaries_of_Phase_Transitions}. 

\bec
\begin{table}[!htp] 
\begin{tabular}{ c  c }
\hline
\hline
Condition & $~~$Intersection \\
 & number \\
 \hline
$\Sigma>0$& $\leq3$ \\

$ \Sigma<0$ & $\leq1$\\
  \hline
  \hline
\end{tabular}
\caption{Constraints on the intersection number for the critical point $\phi_0$ when the action is ${\cal PT}$ symmetric.}
\label{tab:pt-symmetry-phi0}
\end{table}
\eec

\bec
\begin{table}[!htp] 
\begin{tabular}{ c  c c c c}
\hline
\hline
Condition & Intersection \\
 & number \\
 \hline
 $\Delta_R>0$, $\Delta_L>0$, $\Sigma>0$ & $\leq3$ \\
 
 $\Delta_R=0$, $\Delta_L>0$, $\Sigma>0$, $\Pi<0$ & $\leq2$ \\
 $\Delta_R>0$, $\Delta_L=0$, $\Sigma>0$, $\Pi>0$ & $\leq2$ \\
 $\Delta_R<0$, $\Delta_L>0$, $\Sigma>0$, $\Pi>0$ & $\leq2$ \\
 $\Delta_R>0$, $\Delta_L<0$, $\Sigma>0$, $\Pi<0$ & $\leq2$ \\
 
 $\Delta_R=0$, $\Delta_L<0$, $\Sigma>0$ & $\leq1$ \\
 $\Delta_R>0$, $\Delta_L<0$, $\Pi>0$ & $\leq1$ \\
 $\Delta_R>0$, $\Delta_L=0$, $\Sigma<0$, $\Pi>0$ & $\leq1$ \\
 $\Delta_R>0$, $\Delta_L>0$, $\Sigma<0$, $\Pi>$ & $\leq1$ \\
 $\Delta_R<0$, $\Delta_L=0$, $\Sigma>0$ & $\leq1$ \\
 $\Delta_R<0$, $\Delta_L>0$, $\Pi<0$ & $\leq1$ \\
 $\Delta_R=0$, $\Delta_L>0$, $\Sigma<0$, $\Pi<0$ & $\leq1$ \\
 $\Delta_R>0$, $\Delta_L>0$, $\Sigma<0$, $\Pi<0$ & $\leq1$ \\
 
 Otherwise & $=0$ \\
 \hline
 \hline
\end{tabular}
\caption{Constraints on the intersection number for the critical points $\phi_\pm$ when the action is ${\cal PT}$ symmetric.}
\label{tab:pt-symmetry-phipm}
\end{table}
\eec

\bec
\begin{figure*}[!htp]
\subfloat[$\{a,b,c,d,h\}=\{1/5,0,1/5,0,1/100\}$]{\includegraphics[width=6cm]{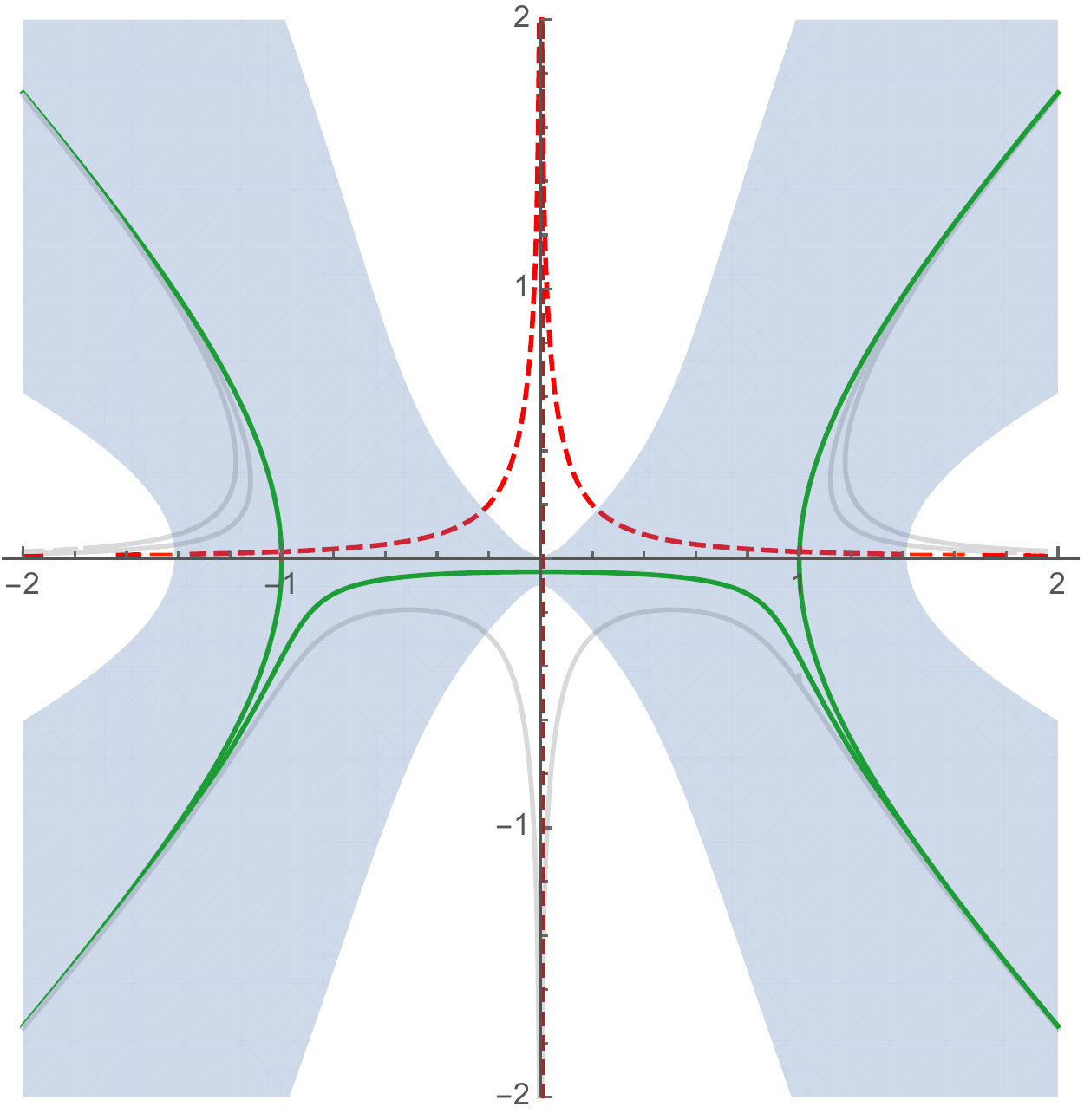}}
$~~~$
\subfloat[$\{a,b,c,d,h\}=\{-1/5,0,1/5,0,1/100\}$]{\includegraphics[width=6cm]{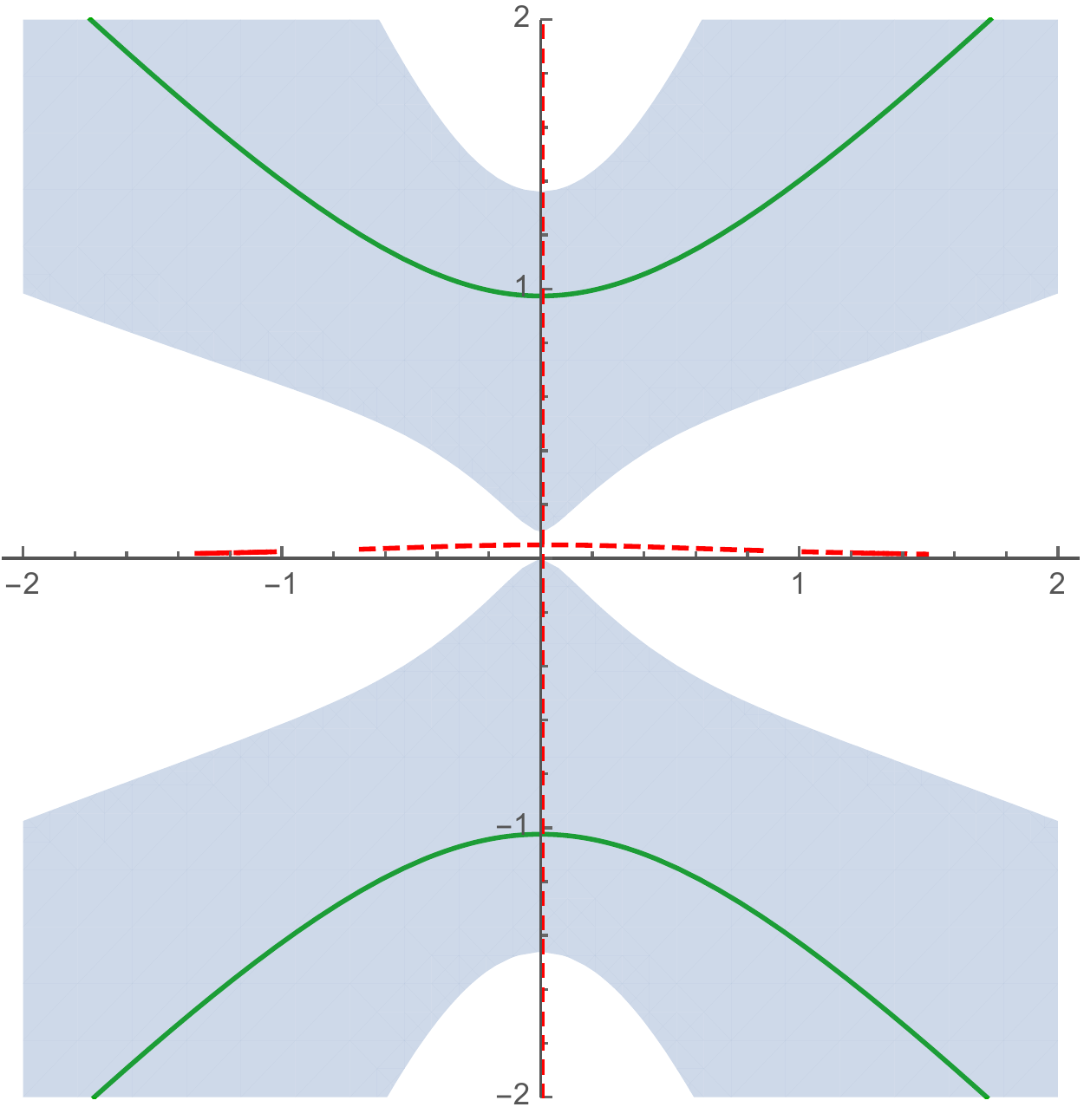}}
\caption{(Colour online). Change in the structure of thimbles as the parameter $a$ crosses the phase boundary $a = 0$. In both the figures, the green solid curves represent the thimbles, red dashed curves represent the anti-thimbles, and the grey solid curves represent the ghosts. The shaded regions represent the regions where $\text{Re} (S) \geq 0$. The anti-thimble $x = 0$ has been offset to $x = 0.01$ for better visibility. We see that there is a drastic change in the underlying thimble structure as the system passes through a phase transition.}
\label{fig:pt-symmetry}
\end{figure*}
\eec

When $a \leq 0$, the situation is far more delicate than the previous situations we have considered. In the region for the these values of the parameters, all the three critical points lie on the imaginary axis ($x = 0$). Further, one of the solutions to the thimble equation, Eq. \eqref{eq:thimble-equation}, is $x = 0$. Since $c < 0$, this solution lies outside the regions of stability, and is an anti-thimble as illustrated in Fig. \ref{fig:pt-symmetry}. The main assumption in deriving Eq. \eqref{eq:int-number-delta} was that the critical points do not share a common gradient flow. This assumption is violated when $a \leq 0$, resulting in the possibility of critical points sharing a common (anti-)thimble, and the (anti-)thimbles of two different critical points intersecting with each other. Thus, in this situation, the intersection number cannot be determined using the method employed in our calculations.  

It would certainly be interesting to explore the intersection numbers and the thimble structures in ${\cal PT}$-symmetric theories in higher dimensions.

\section{Quantum Phase Transition and Change in Thimble Structure}
\label{sec:Phase_Transition_Boundaries_Examples}

In this section, we demonstrate the usefulness of the results in Sec. \ref{sec:Phase_Transition_Boundaries} with the help of a few examples. 

We choose to fix the parameters $c$, $d$, and $h$, and vary either $a$ or $b$ in order to maximize the number of conditions that need to be checked. 

\begin{figure}[!htp]
\subfloat[The partition function]{\includegraphics[width=8cm]{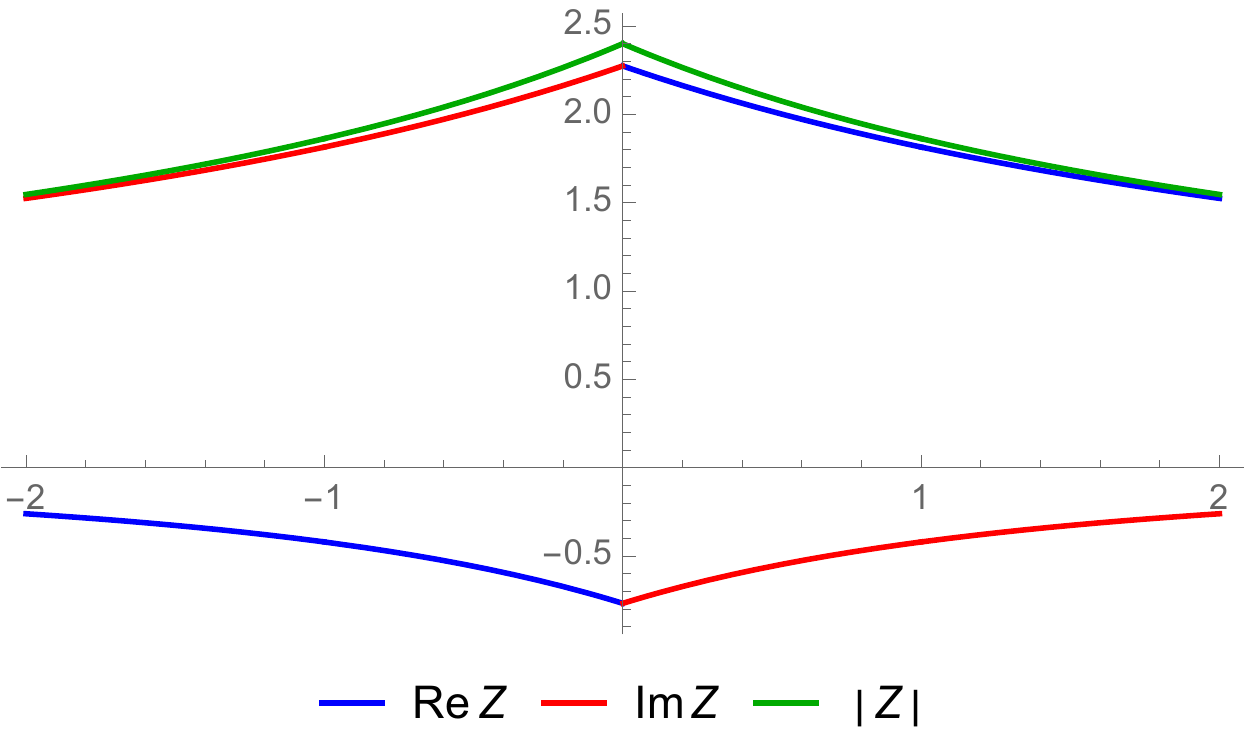}}
$~~~$
\subfloat[The observable $\langle\phi^{2}\rangle$]{\includegraphics[width=8cm]{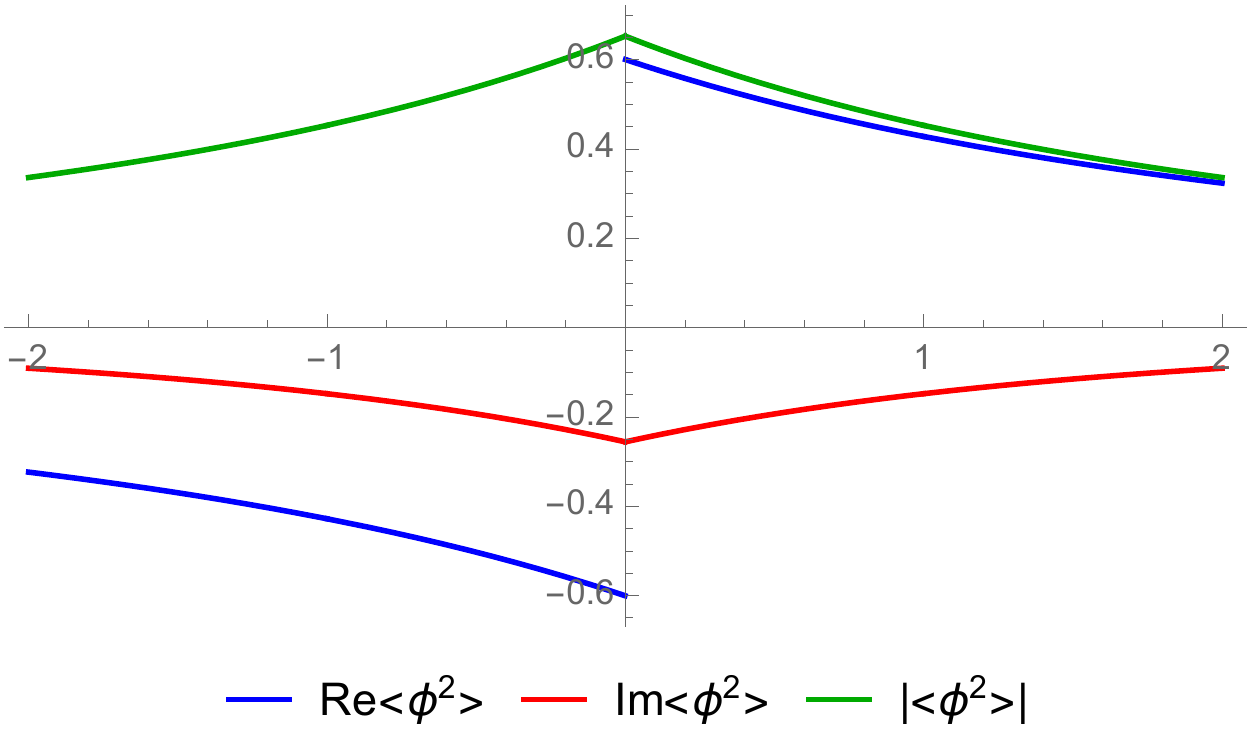}}
\caption{(Colour online). The partition function $Z \equiv Z(a, b, c, d, h)$ and the observable $\langle \phi^2 \rangle$ as a function of $a$ for the fixed parameters $\{b = 1, c = 1, d = 0, h = 0 \}$. The blue curve represents the real part, the red curve represents the imaginary part, and the green curve represents the absolute value. Clearly, there is a discontinuity/kink at $a = a_c = 0$. This discontinuity is due to the change in the intersection number of $\phi_{\pm}$ from zero, for $a > 0$, and one for each critical point, when $a < 0$.}
\label{fig:z_a_1_1_0_0}
\end{figure}

\begin{figure*}[!htp]
\subfloat[$\{a,b,c,d,h\}=\{1,1,1,0,0\}$]{\includegraphics[width=6cm]{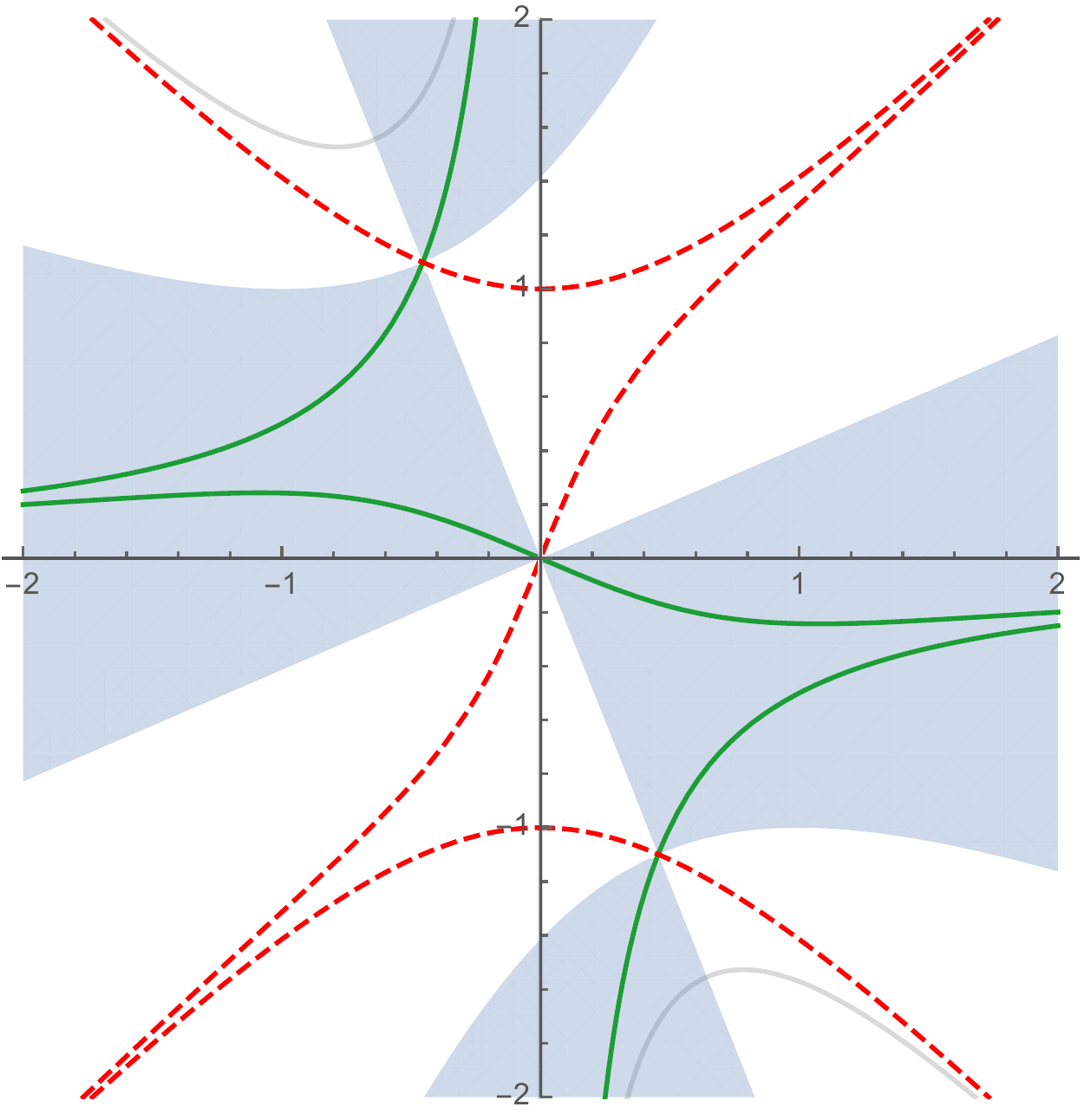}}
$~~~$
\subfloat[$\{a,b,c,d,h\}=\{-1,1,1,0,0\}$]{\includegraphics[width=6cm]{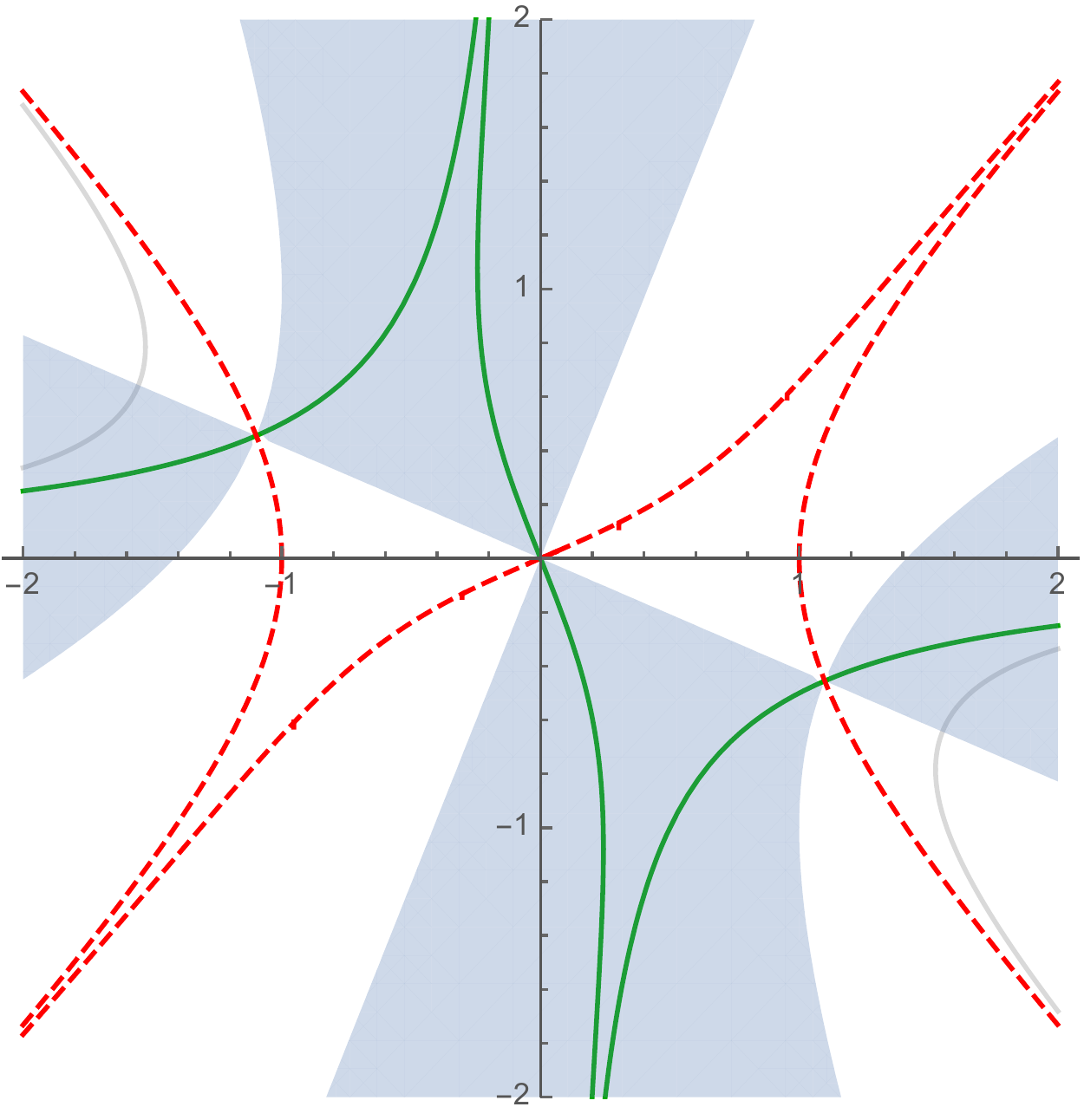}}
\caption{(Colour online). Change in the structure of thimbles as the parameter $a$ crosses the phase boundary $a = a_c =0$. In both the figures, the green solid curves represent the thimbles, red dashed curves represent the anti-thimbles, and the grey solid curves represent the ghosts. The shaded regions represent the regions where $\text{Re}(S) \geq 0$. We see that there is a drastic change in the underlying thimble structure as the system passes through a phase transition.}
\label{fig:topology-change}
\end{figure*}

First, consider the situation where $\lambda = 1$ and $h = 0$. Equation \eqref{eq:ni-d=0} tells us that the intersection number depends only on the relative sign of $a$ and $c$, and that $b$ has no effect on the intersection number. Thus, choosing $b = 1$ and looking at the the partition function as a function of $a$, we clearly observe a discontinuity/kink at the critical value of the parameter $a = a_c = 0$. (We can see that this discontinuity is a result of branch cut crossing in the $\sigma$ plane.) It is shown in in Fig. \ref{fig:z_a_1_1_0_0}. Thus for the given choice of parameters, the system undergoes a phase transition at $a = 0$. Looking at the corresponding change to the structure of the thimbles, shown in Fig. \ref{fig:topology-change}, the discontinuity in $Z$ is due to the change in the intersection number of $\phi_{\pm}$ from zero, for $a > 0$, and one for each critical point, when $a < 0$.

\bec
\eec

\bec
\eec

We now choose to vary $b$ after fixing the parameter $\{a = 1, c = 1, d = 1, h = 0\}$. The expressions in Eq. \eqref{eq:exp-for-cond-h=0-dneq0} take the following forms in terms of $b$
\beq
\Delta = \frac{(b-1)^2}{2}, ~~ \Pi = \frac{b^2 + 2b - 1}{2}, ~~ \Sigma = b.
\eeq

\begin{figure}[!htp]
\subfloat[The partition function]{\includegraphics[width=8cm]{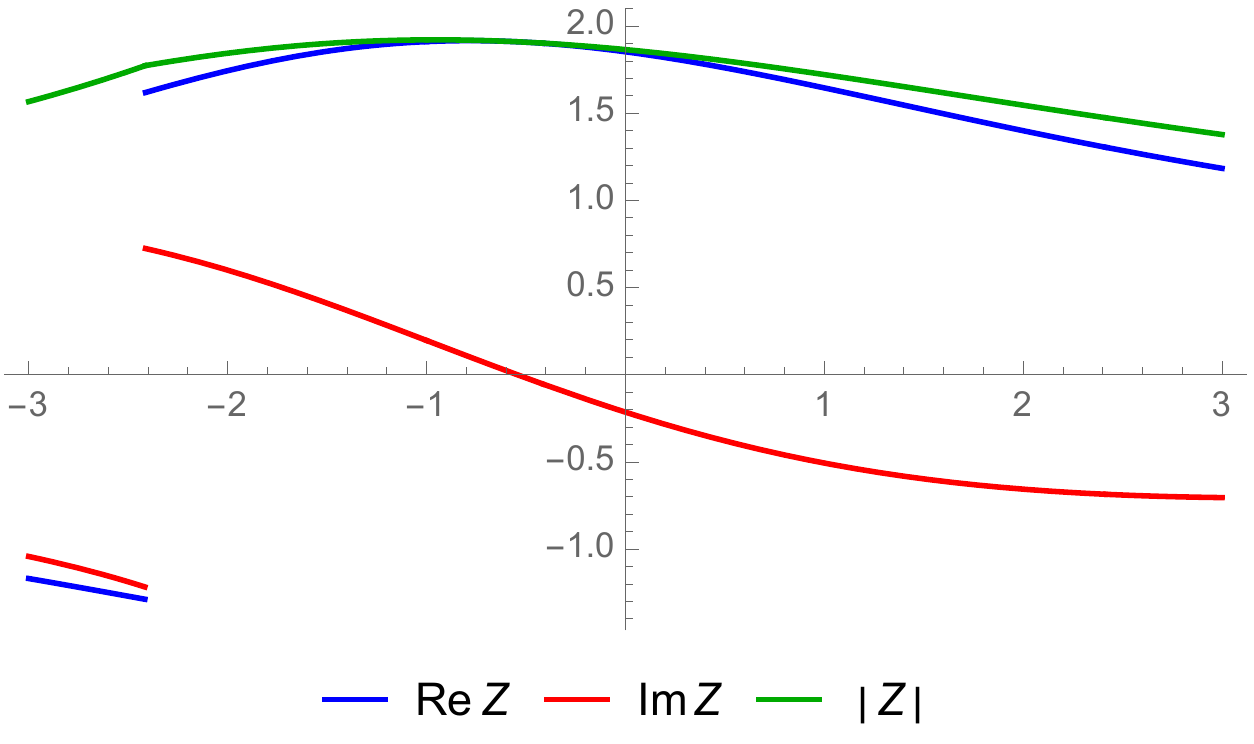}}
$~~~$
\subfloat[The observable $\langle\phi^{2}\rangle$]{\includegraphics[width=8cm]{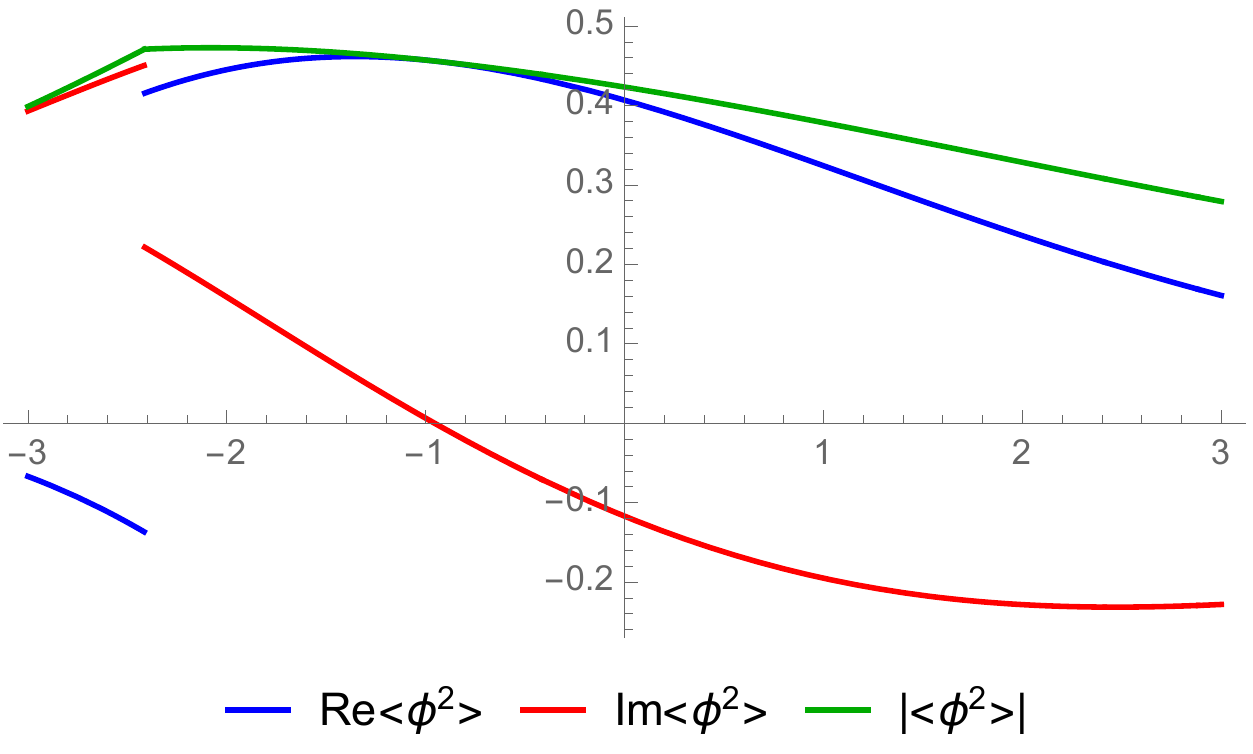}}
\caption{(Colour online). The partition function $Z$ and observable $\langle \phi^2 \rangle$ as a function of $b$ for the parameters $\{a = 1$, $c = 1$, $d = 1$, $h = 0\}$. The blue curve represents the real part, the red curve represents the imaginary part, and the green curve represents the absolute value. We observe a discontinuity at $b = b_c = -1 - \sqrt{2}$.}
\label{fig:z_1_b_1_1_0}
\end{figure}

Based on the conditions given in Table \ref{tab:hzero-dnotzero-phi0}, we expect a sudden change in the value of the partition function at the critical coupling $b = b_c = 0$. From Table \ref{tab:hzero-dnotzero-phipm}, we expect that this should happen when $b = 0, -1 - \sqrt{2}$. Note that although it seems like we can expect a phase transition around $b = 1$ and $b = -1 + \sqrt{2}$, in the vicinity of these points, the intersection number does not change. On plotting the partition function for these parameters, we observe a discontinuity at $b = -1 - \sqrt{2}$. (See Fig. \ref{fig:z_1_b_1_1_0}.) The explanation for why we do not obtain a discontinuity is that at $b = 0$, the change in the number of solutions is reflected in $\langle{\cal J}_i, \mathbb{R}\rangle$ instead of $\langle{\cal K}_i, \mathbb{R}\rangle$. This explains why we have mentioned everywhere that the intersection number is less than or equal to a certain integer.

\bec
\eec

Let us look at a slightly more complicated case. We can try to find discontinuities as we vary both the couplings $a$ and $b$ simultaneously. The expressions given in Eq. \eqref{eq:exp-for-cond-h=0-dneq0} in terms of $a$ and $b$ for $\{ c = 1, d = 1, h = 0 \}$ are
\beq
\Delta = \frac{(b - a)^2}{8}, ~~ \Pi = \frac{b^2 + 2ab - a^2}{2}, ~~ \Sigma = b.
\eeq

\bec
\begin{figure*}

\subfloat[Real part of the partition function.]{\includegraphics[width=5cm]{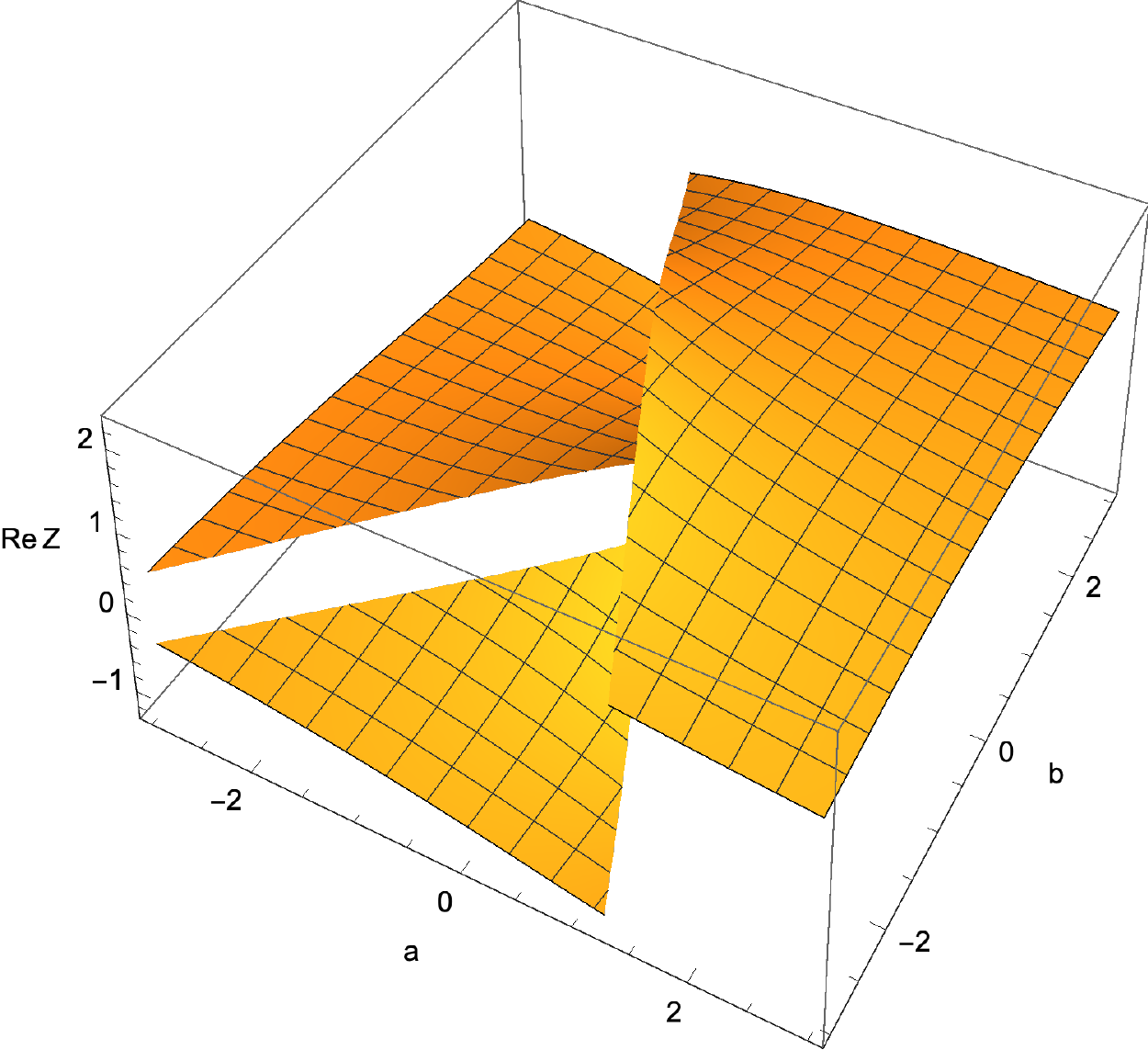}}\label{Z_a_b_wocont_re}
$~~~$
\subfloat[Imaginary part of the partition function.]{\includegraphics[width=5cm]{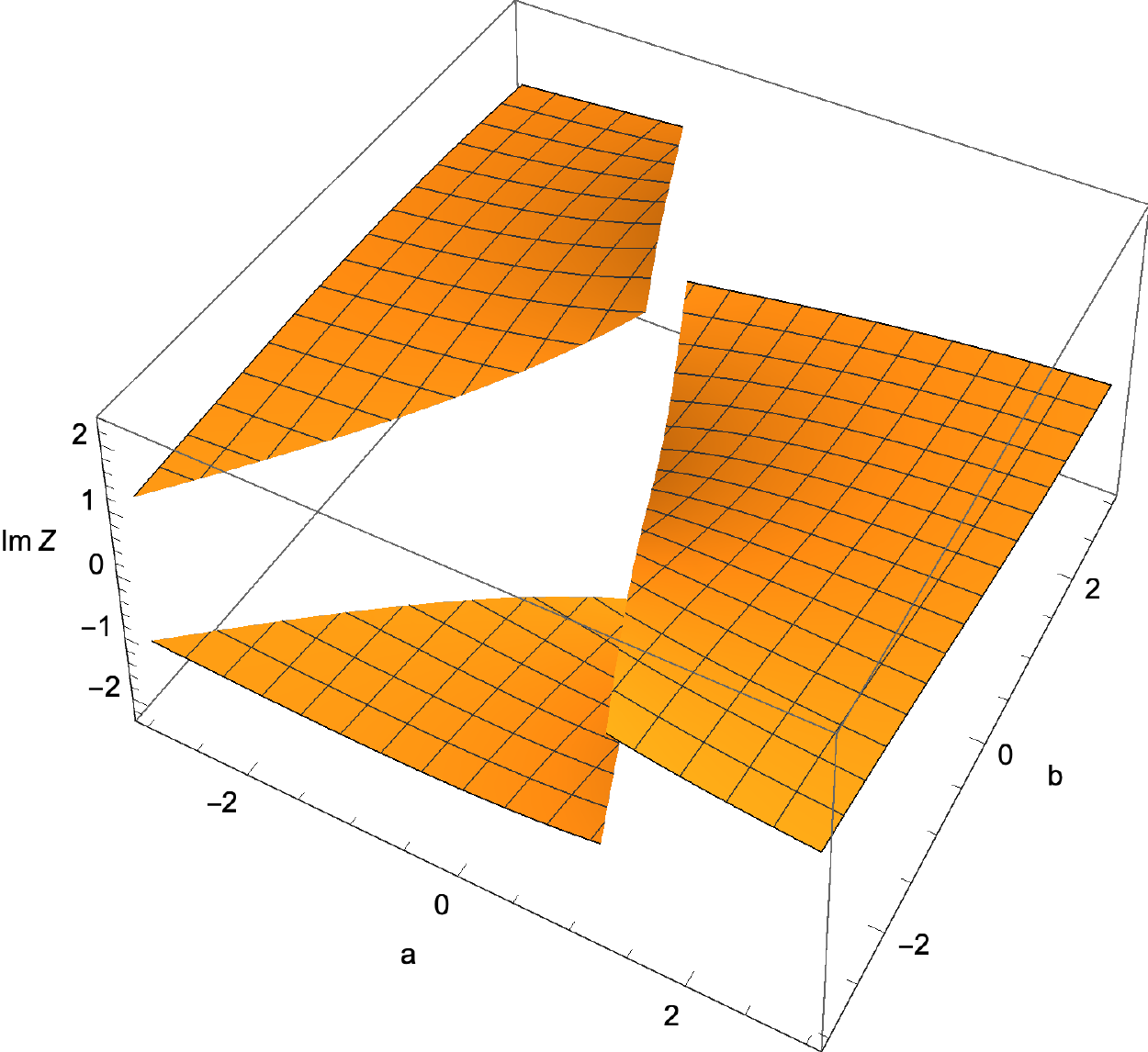}}\label{Z_a_b_wocont_im}
$~~~$
\subfloat[Absolute value of the partition function.]{\includegraphics[width=5cm]{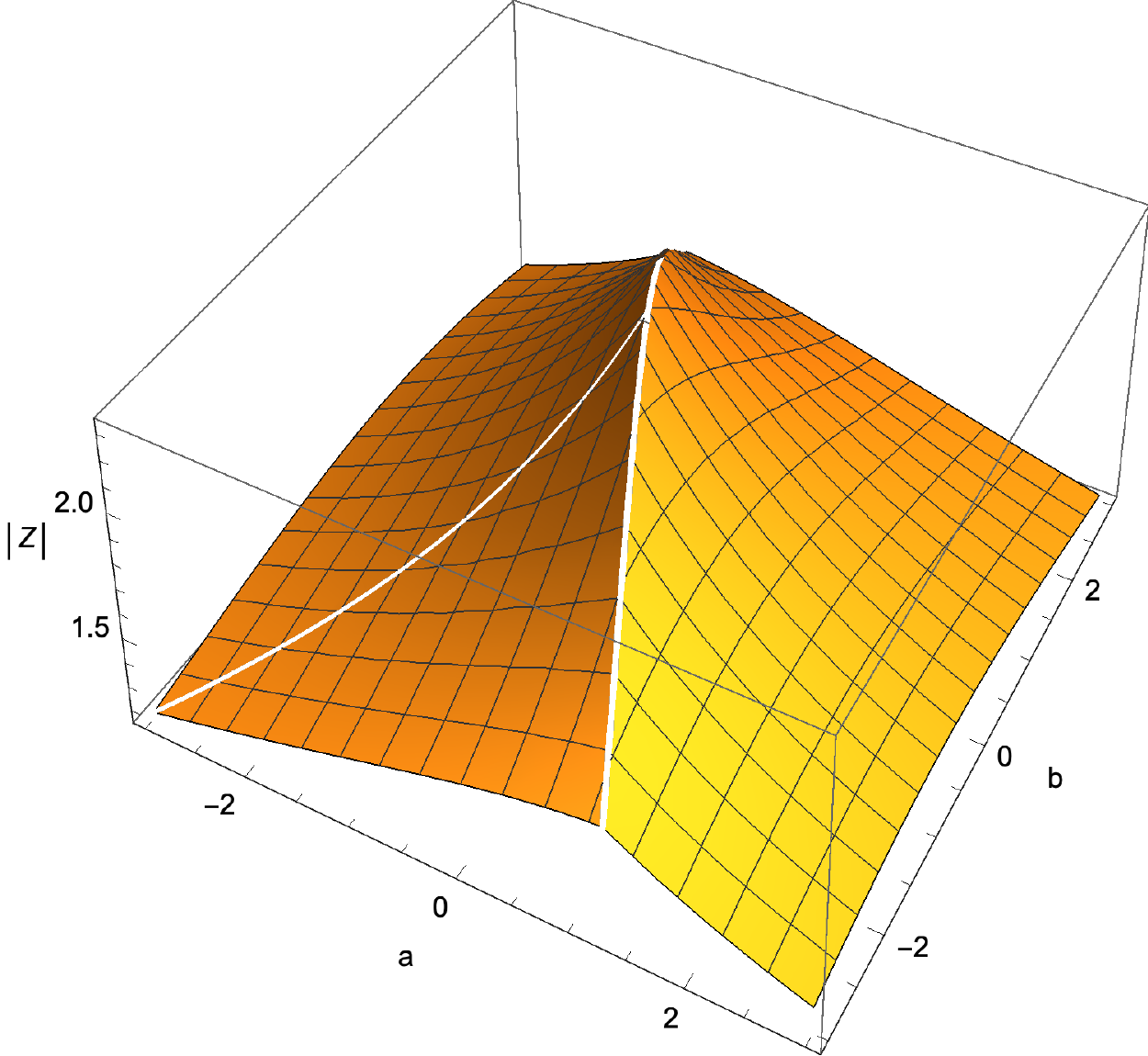}}\label{Z_a_b_wocont_abs}

\subfloat[Real part of the partition function, with the surfaces on which either $a = b$ or $b^2 - a^2 + 2ab = 0$.]{\includegraphics[width=5cm]{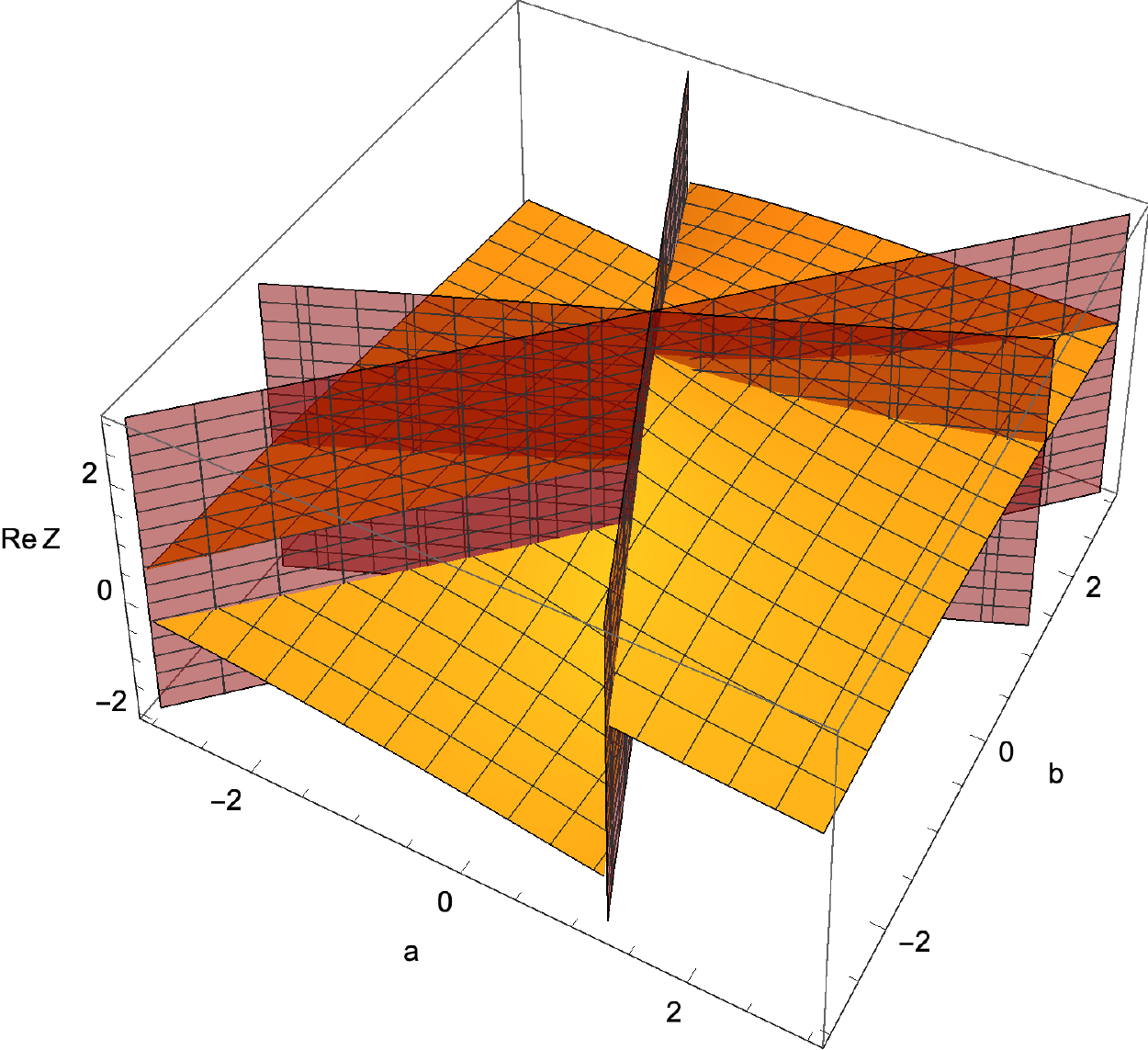}}\label{Z_a_b_wcont_re}
$~~~$
\subfloat[Imaginary part of the partition function, with the surfaces on which either $a = b$ or $b^2 - a^2 + 2ab = 0$.]{\includegraphics[width=5cm]{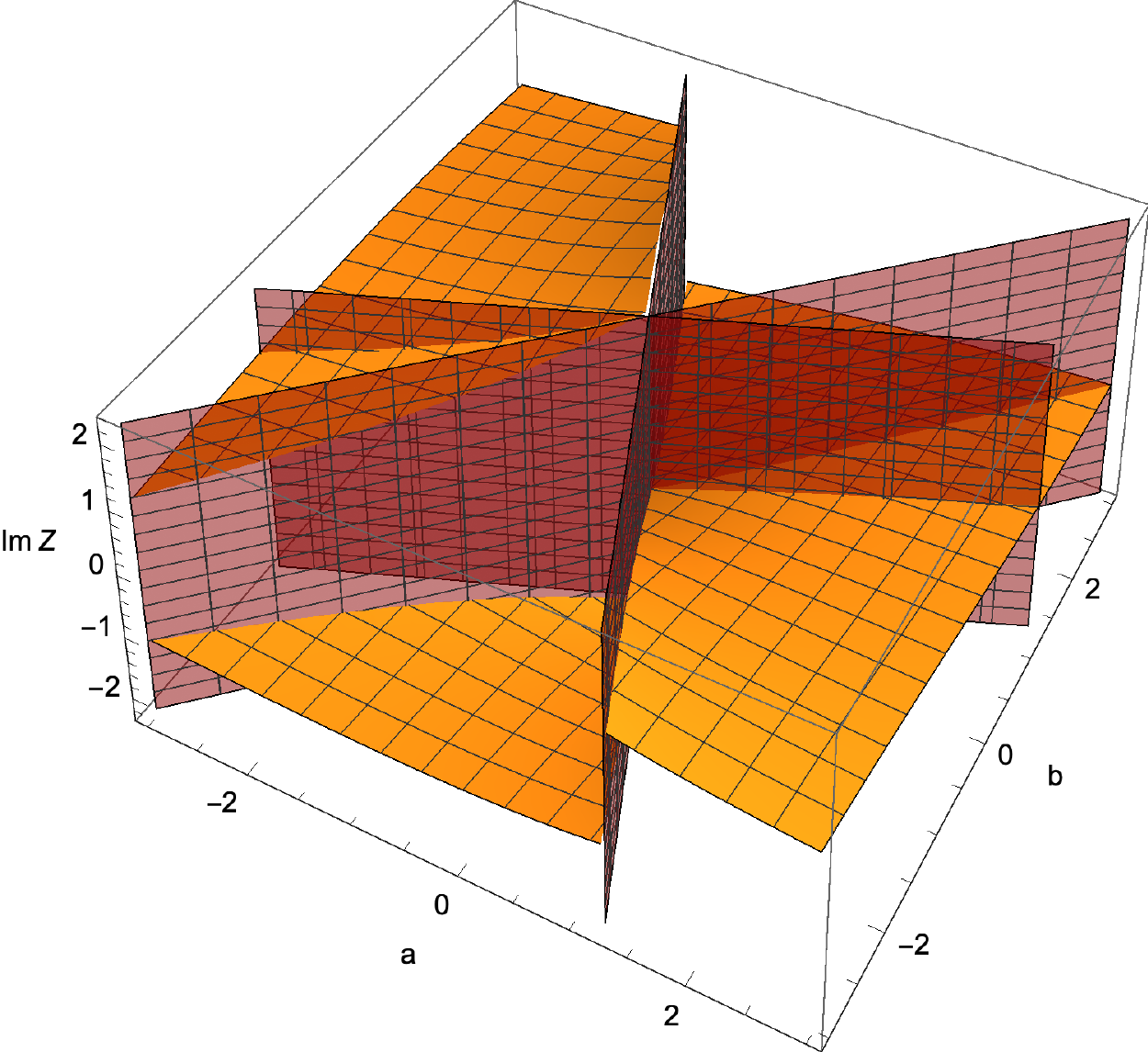}}\label{Z_a_b_wcont_im}
$~~~$
\subfloat[Absolute value of the partition function, with the surfaces on which either $a=b$ or $b^2-a^2+2ab=0$.]{\includegraphics[width=5cm]{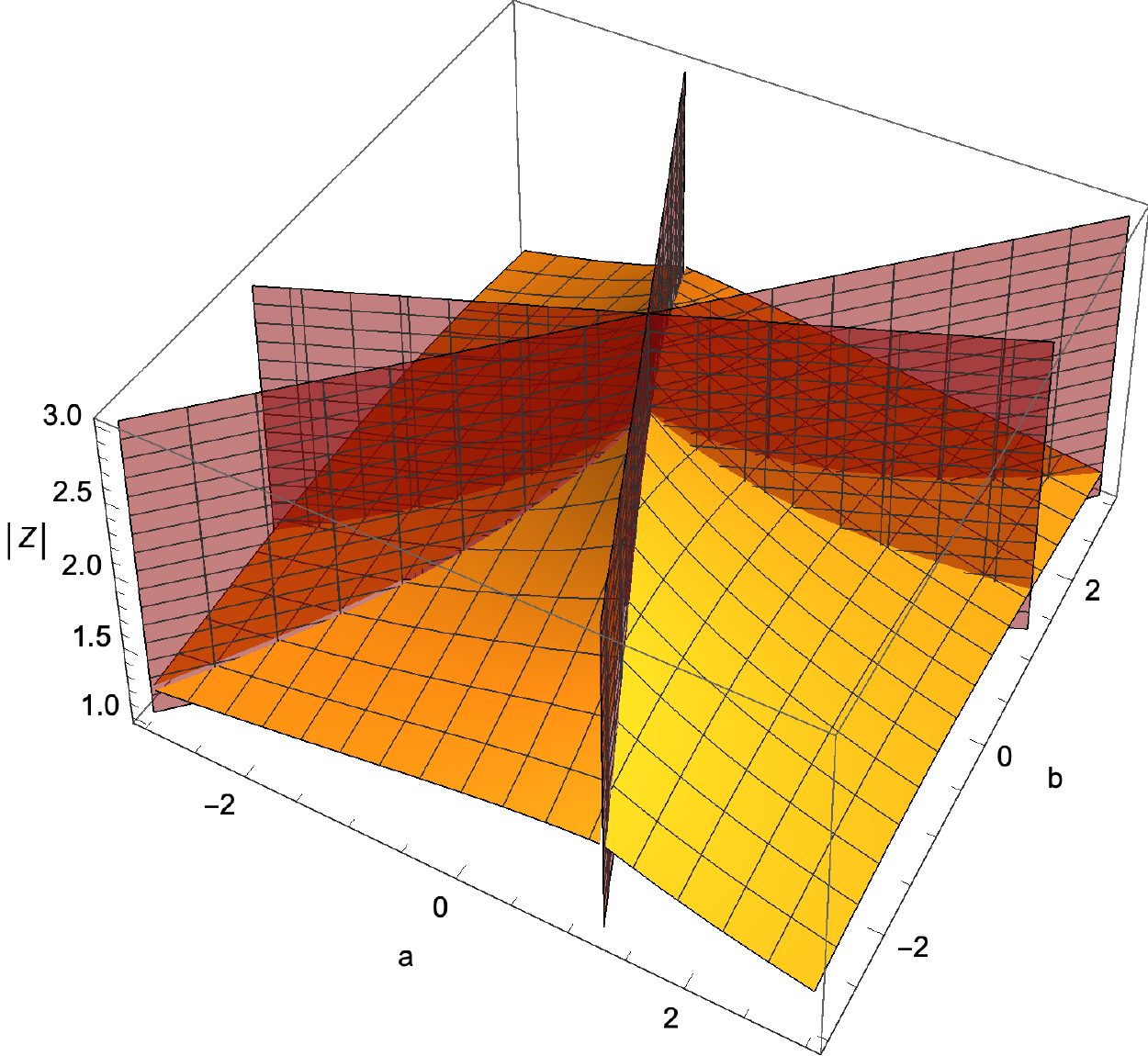}}\label{Z_a_b_wcont_abs}
\caption{\label{fig:3d_plot_Z}(Colour online) Real, imaginary, and absolute values of the partition function as a function of $a$ and $b$ for the parameters $\{c = 1$, $d = 1$, $h = 0\}$. The yellow surfaces represent the respective values of the partition function and the red surfaces represent the surface along which either $a = b$ or $b^2 - a^2 + 2ab = 0$.}
\end{figure*}
\eec

A naive expectation would thus straightforwardly be that when $a = b$ or when $b^2 - a^2 + 2ab = 0$ (corresponding to the case $\Delta = 0$ and $\Pi = 0$), the partition function will have a discontinuity. (See Fig. \ref{fig:3d_plot_Z}.) Plotting the partition function as a function of $a$ and $b$, we observe that this expectation is valid in certain cases, and in certain cases there is no discontinuity. 

Let us now turn on the source term. We maintain $\{a = 1, c = 1, d = 1\}$. Choosing $h = 0.01$, Eq. \eqref{eq:imag-part-action-i} becomes
\beq
\rho_0 = 0, ~~ \rho_+ = \frac{1 - b^2 - 2b}{8} + 0.01 \Omega, ~~ \rho_- = \frac{1 - b^2 - 2b}{8} - 0.01 \Omega,
\eeq
where 
\beq
\Omega = \sqrt{b^2 + 1} \cos\left[\arctan\left(\frac{b - 1}{b + 1}\right) \right].
\eeq
We do not expect a phase transition with respect to $\phi_0$ since here $\rho_0 = 0$, and we have already fixed our choice of $a$ and $c$. Corresponding to the critical points $\phi_+$ and $\phi_-$, we have, from Eq. \eqref{eq:hreal-conditions}
\bea
&& \Delta_+ = \frac{b^2 - 2b + 1}{2} + 0.04 \Omega, ~~ \Delta_- = \frac{b^2 - 2b + 1}{2} - 0.04 \Omega, \nn \\
&& \Pi_+ = \frac{1 - b^2 - 2b}{8} + 0.01 \Omega, ~~ \Pi_- = \frac{1 - b^2 - 2b}{8} - 0.01 \Omega, ~~ \Sigma = b. \nn
\eea

As is evident, when the source term is real, the equivalence between the critical points $\phi_+$ and $\phi_-$ gets lifted while $\phi_0$ remains untouched. Solving the equations using a symbol interpreter, we get the points where the phase transitions could be expected as
\beq
b =
\left\{
\begin{array}{ll}
      0 & {\rm (if~} \Sigma = 0), \\
      1 + \frac{1}{(25\sqrt{2})} \pm {1}{2} \sqrt{\frac{2}{625} + \frac{8 \sqrt{2}}{25}} & {\rm (if~} \Delta_- = 0), \\
      -1 \pm \frac{1}{25}\sqrt{1251 - \sqrt{2501}} & {\rm (if~} \Pi_+ = 0), \\
      -1 \pm \frac{1}{25}\sqrt{1251 + \sqrt{2501}} & {\rm (if~} \Pi_- = 0), \\
\end{array} 
\right. 
\eeq

Numericising the above values of $b$, we get
\beq
b = \left\{ 0, ~0.6907, ~1.3658, ~-2.3862, ~0.3862, ~-2.4428, ~0.4428 \right\}.
\eeq

The partition function for this action, from Eq. \eqref{eq:partition-function-sources}, is given as
\beq
Z = I_0 + \frac{h^2}{2} I_2 + {\cal O}(h^4) = I_0 + 0.00005I_2.
\eeq

This has a discontinuity in the vicinity of $b = -2.4$. However, the point at which the partition function is discontinuous does not match exactly with either $b = -2.38621$ or $b = -2.44278$. In fact, it matches exactly with our previous example where the boundary was at $b = -1 - \sqrt{2}$. We believe the issue is with the expansion of $Z_{\text{sources}}$ and not the method used to find the points of phase transitions, due to the fact that the perturbative expansion with respect to $h$ in Eq. \eqref{eq:partition-function-sources} depends on the partition function and observables of the action without sources. These are not sensitive to the lifting of equivalence between $\phi_+$ and $\phi_-$.

\section{Summary of Results}
\label{sec:Summary_of_Results}

In this paper, we have considered a zero-dimensional scalar field theory with quartic interactions and a source term - a model that captures the simplest nontrivial quantum field theory action. In this theory, the thimbles can be found analytically by exploiting the most crucial property of these curves -- the imaginary part of the action remains constant on them. However, solving for the thimbles using this method has its own problems as illustrated in Sec. \ref{sec:Thimble_Equations_and_Observables} where for more general situations, it is difficult to clearly distinguish between the solutions as they can either be thimbles or anti-thimbles, based on the region in the complex plane under question. We called this the `piecewise behavior' of the solutions. To our knowledge, the piecewise behavior is not unique to the model we have considered. It would be interesting to comment about the piecewise behavior of thimbles when dealing with models in higher dimensions. 

Despite these issues, there are advantages of employing the Lefschetz thimbles method since it provides a lot of ancillary information about the system. Since the intersection numbers in Eq. \eqref{eq:z-integral} are in general integers, changes in the intersection numbers correspond to discontinuities in the partition function and observables, indicating the existence of different phases, characterized by the coupling parameters of the model. We used the simple method of solving Eq. \eqref{eq:thimble-equation}, massaged in a way to access the information on the intersection numbers as outlined in Sec. \ref{sec:Phase_Transition_Boundaries} and Appendix \ref{sec:Expressions_for_Boundaries_of_Phase_Transitions}, to find conditional expressions involving relations between the coupling parameters of the system that characterize the different phases. A few examples showcasing the effectiveness of this method was presented in Sec. \ref{sec:Phase_Transition_Boundaries_Examples}.

Although from the results, it is evident that quantum phase transitions occur in the system as the control parameter is varied through the critical point, we note that the theory exists at zero temperature. The parameters that tell us about the phase transitions in the model are non-thermal parameters and this is the reason these phase transitions are called quantum phase transitions or quantum critical points \cite{Cherman:2014ofa, Guralnik:2007rx,Kanazawa:2014qma,Basar:2013eka}. There are two main observations about the behavior of the phases. First, the boundaries of phase transitions are completely determined by the parameters $\sigma, \lambda$, and $h$. Thus, any symmetry involving the field $\phi$ would remain a symmetry after the phase transition. Second, these phase transition boundaries correspond to distinct changes in the topological structure of the thimbles and anti-thimbles. (We show this feature in Fig. \ref{fig:topology-change}.) Further, regions within the phase boundaries are akin to {\it wall chambers} and phase transitions correspond to {\it wall crossing} \cite{Witten:2010cx}.

We also note that comments on the thermodynamical nature of these transitions cannot be made for the model we have chosen since thermodynamic quantities such as the free energy cannot be consistently defined in zero dimensions. If we study this model in one or more dimensions we could talk about the interplay between quantum and thermal phase transitions. We leave the investigation of the thimble structure and quantum/thermal phase transitions of the next nontrivial system, a one-dimensional model defined on a Euclidean thermal circle, for the future.

\section{Conclusions and Future Directions}
\label{sec:Conclusions_and_Future_Directions}

In this paper, we exploited the properties of Lefschetz thimbles to analytically demonstrate how the thimble formalism can be used to predict quantum critical points in nontrivial zero-dimensional scalar field theories. An immediate extension would be to explore the same problem for zero-dimensional supersymmetric quantum field theories. A supersymmetric version of the zero-dimensional $\mathcal{PT}$-symmetric model was recently studied using complex Langevin dynamics in Ref. \cite{Joseph:2019sof}. 

In zero-spacetime dimensions, except for showing that the partition function and observables develop discontinuities as one or more non-thermal parameters are varied, leading to quantum phase transitions in the system, comments on the thermodynamical nature of phase transitions cannot be made. Thus a more nontrivial and highly elucidatory extension would be to study phase transitions in higher dimensional systems, where the information of the background manifold becomes important and thermodynamic quantities can be defined. It would be interesting to study the interplay between quantum critical points and thermodynamic critical points when the higher dimensional quantum field theories in question are put on a Euclidean thermal circle.

There has been some success in effecting these calculations numerically using hybrid Monte Carlo simulations for the one-dimensional Thirring model \cite{Fujii:2015vha}, and there are numerous demonstrations of connections between Lee-Yang zeroes and Stokes phenomena in the context of chiral phase transitions \cite{Kanazawa:2014qma, Guralnik:2007rx, Itzykson:1983gb, pisani_lee-yang_1993}. However, a completely analytic and general demonstration of phase structures of higher dimensional systems, their relation to the structure of thimbles/anti-thimbles, and a relation with the thermodynamics of the system if any, is desired.

Another nontrivial and interesting generalization would be introducing non-Abelian degrees of freedom in the zero-dimensional theory. The IKKT matrix model, a zero-dimensional supersymmetric non-Abelian quantum field theory, serves as a promising candidate for a nonperturbative formulation of superstring theory. However, this model is shown to have a complex fermion operator \cite{Anagnostopoulos:2020xai, Aoki:2019tby, Nishimura:2019qal} making the effective bosonic action complex. An investigation based on the thimble formalism would certainly turn out to be fruitful.

We note that in the recent past, complexification methods in the study of path integrals have been the focus of quite a bit of analytic and numerical work. The immediate hope would be generalizing the analysis to higher-dimensional integrals, which would either be obtained by putting the path integral on a lattice, or due to internal degrees of freedom when the fields are taken to be matrix-valued for a system in zero dimensions. Although the task would be formidable, we believe that it should not be unachievable. In the case of putting the path integral on a lattice, the mass and the interaction terms of the model only involve the field $\phi_n$ defined at a lattice site $n$ and would not add to the difficulty of solving the problem. The parts that could lead to difficulties are the hopping terms, such as $\phi_n\phi_{n+1}$ . That, however, should not be extremely difficult since the method we prescribe talks about existence and number of sections of hypersurfaces described by multi-variate polynomials of order less than five.

There exists a construction that automatically finds the right combination of thimbles, without the need to compute any intersection numbers, and has already been shown to produce physically interesting results in Monte Carlo simulations of a variety nontrivial quantum field theories \cite{Alexandru:2015xva, Alexandru:2016san, Alexandru:2018ngw, Alexandru:2017lqr, Alexandru:2016lsn, Alexandru:2016ejd, Alexandru:2016gsd, Alexandru:2017oyw, Alexandru:2017czx, Alexandru:2018fqp, Alexandru:2018ddf}. Our work, though perhaps is useful primarily for analytic work, provides an alternative method to finding the relevant thimbles for the choice of parameters that have been made. Though it may be doubtful whether it would be useful for numerics, it would certainly be useful for analytic work.

\acknowledgments 

The work of AJ was supported in part by the Start-up Research Grant (No. SRG/2019/002035) from the Science and Engineering Research Board (SERB), Government of India, and in part by a Seed Grant from the Indian Institute of Science Education and Research (IISER) Mohali. The work of RB was partially supported by an INSPIRE Scholarship for Higher Education by the Department of Science and Technology, Government of India. A part of this work was presented by RB at the Science Undergraduate Research Conference at Azim Premji University, Bangalore, India in December 2019.

\appendix

\section{Expressions for Boundaries of Phase Transitions}
\label{sec:Expressions_for_Boundaries_of_Phase_Transitions}

In this section we derive the expressions for the boundaries of phase transitions. We start with the case where the source parameter $h$ is zero. The imaginary part of the action is 
\beq
\label{eq:imag-part-action-app}
{\rm Im} ~S(x,y) = \frac{d}{4} y^4 - cxy^3 - \left( \frac{b}{2} + \frac{3d}{2} x^2 \right) y^2 + \left( ax + cx^3 \right) y + \frac{b}{2} x^2 + \frac{d}{4} x^4,
\eeq
and the critical points of the action are 
\beq
\phi_0 = 0, ~~\phi_{\pm} = \pm i \sqrt{\frac{\sigma}{\lambda}}.
\eeq

Along the original integration cycle, $\mathbb{R}$, which corresponds to $y = 0$, the imaginary part of the action is 
\beq
\label{eq:action-along-R}
{\rm Im} ~S(x) = \frac{b}{2} x^2 + \frac{d}{4}x^4.
\eeq

To get the combined intersection number of the thimble and anti-thimble of a critical point, we look for the existence of real solutions to the equation
\beq
\label{eq:action-along-R-critical-point}
\frac{b}{2} x^2 + \frac{d}{4} x^4 - \rho_i = 0.
\eeq

Here, $\rho_i = {\rm Im~} S$ at $\phi_i$, see Eq. \eqref{eq:imag-part-action-i}. In the case where $d = 0$, Eq. \eqref{eq:action-along-R-critical-point} is simply
\beq
x = \pm \sqrt{\frac{2 \rho_i}{b}}.
\eeq

Substituting $\rho_i$, the above equation takes the form
\beq
x = \left\{
\begin{array}{ll}
      0 & {\rm ~~~~when~} i = 0, \\
      \pm \sqrt{-\frac{a}{c}} & {\rm ~~~~when~} i = \pm. \\
\end{array} 
\right. 
\eeq

Requiring $x \in \mathbb{R}$ and using $\left \langle \mathcal{J}_i, \mathcal{K}_{i'} \right \rangle = \delta_{i, i'}$, this immediately gives us
\beq
n_i \left\{
\begin{array}{ll}
     = 1 & {\rm ~~~~when~} i = 0, \forall ~~ \alpha, \beta, \\
     \leq 1 & {\rm ~~~~when~} i = \pm, \frac{a}{c} < 0, \\
     = 0 & {\rm ~~~~when~} i = \pm, \frac{a}{c} > 0. \\
\end{array} 
\right. 
\eeq

When $d \neq 0$, Eq. \eqref{eq:action-along-R-critical-point} is a bi-quadratic. Defining $w = x^2$, we look for \emph{positive} real solutions to
\beq
\frac{b}{2} w + \frac{d}{4} w^2 - \rho_i = 0.
\eeq

The relevant parameters associated with the above equation are its discriminant, sum of roots, and product of roots. Denoting them as $\Delta$, $-\Sigma$, and $\Pi$ respectively\footnote{We use $-\Sigma$ for sum of roots to avoid dealing with an overall minus sign in our results.}, we obtain
\beq
\Delta = \frac{(bc - ad)^2}{(c^2 + d^2)}, ~~\Pi = \frac{d (b^2 - a^2) + 2abc}{d (c^2 + d^2)}, ~~\Sigma = \frac{b}{d}.
\eeq

It is to be noted that we have chosen to omit overall factors (such as that of $2$ in $\Sigma$) since what is relevant is only the sign of these quantities. When the discriminant is positive, we have two distinct real solutions for $w$. In this case, when the product of roots is positive, either both solutions are positive (giving a combined intersection number of 4) or both solutions are negative (intersection number is zero). This is checked using $\Sigma$. When the discriminant is zero, we only get one real root for $w$. Again, $\Sigma$ helps in determining whether the root is positive or negative. When the discriminant is negative (which for this particular case is never possible), there are no real roots of $w$ and the intersection number is zero. These end up giving the conditions mentioned in Tables \ref{tab:hzero-dnotzero-phi0} and \ref{tab:hzero-dnotzero-phipm}. 

When the source term is non-zero, and the parameter $h$ is real, the analysis remains exactly the same. The only change is the change to $\rho_i$. If the source term is purely imaginary, Eq. \eqref{eq:action-along-R} now becomes 
\beq
hx + \frac{b}{2}x^2 + \frac{d}{4}x^4 + \rho_i = 0. 
\eeq

Since the equation is now a purely quartic equation (in the sense that it is not reducible to a bi-quadratic), we have a complicated set of conditions for $n_i$. We refer the reader to the conditions in Ref. \cite{rees_graphical_1922} to arrive at the results in Table \ref{tab:himag}.


When the action possesses ${\cal PT}$ symmetry, we can obtain its critical points and the imaginary part of the action by substituting $h \rightarrow ih$ and $c \rightarrow -c$ in Eqs. \eqref{eq:critical-points} and \eqref{eq:imag-part-action-i}, which gives us
\bea
\phi_0 &=& -\frac{ih}{a} + \mathcal{O}(h^3), \\ 
\phi_\pm &=& \pm \sqrt{\frac{a}{c}} - \frac{ih}{2a}\pm \frac{3h^2}{8}\sqrt{\frac{c}{a^5}} + \mathcal{O}(h^3).
\eea

\beq
\rho_i = \left\{
\begin{array}{ll}
      0 & {\rm when~}  a \leq 0, ~\forall ~i, \\
      0 & {\rm when~}  a > 0, ~i = 0, \\
      - h \sqrt{\frac{a}{|c|}} & {\rm when~}  a > 0, ~i = +, \\
      + h \sqrt{\frac{a}{|c|}} & {\rm when~}  a > 0, ~i = -. \\
\end{array} 
\right. 
\eeq

The imaginary part of the action, given in Eq. \eqref{eq:imag-part-action-app}, upon making these substitutions becomes
\beq
\label{eq:imag-part-action-pt-app}
{\rm Im} ~S(x,y) = hx + axy - c \left( x^3y - xy^3 \right ). 
\eeq

As outlined in Sec. \ref{sec:Quartic_Model_with_a_Source_Term}, the standard procedure for dealing with this action is to take an integration cycle about the angles $5\pi/4$ and $7\pi/4$ (in the third and the fourth quadrant, respectively). We have chosen it to be (See Eq. \eqref{eq:pt-symmetry-cycle})
\beq
y(x) = \left\{
\begin{array}{ll}
      x & {\rm for~} x\leq 0, \\
      -x & {\rm for~} x > 0. \\
\end{array} 
\right. 
\eeq

Substituting the above integration cycle in Eq. \eqref{eq:imag-part-action-pt-app} and equating it to the imaginary part of the action at a critical point, we obtain
\bea
\label{eq:left-thimbles}
hx + ax^2 + \rho_i &=& 0 ~~ {\rm for~} x\leq 0, \\
\label{eq:right-thimbles}
hx - ax^2 + \rho_i &=& 0 ~~ {\rm for~} x > 0.
\eea

\bec
\begin{table}[!htp] 
\begin{tabular}{ c  c }
\hline
\hline 
Condition & Intersection \\
 & number \\
 \hline
 $\Delta_L>0$, $\Pi_L>0$, $\Sigma_L<0$ & $\leq2$ \\
 $\Delta_L=0$, $\Sigma_L<0$ & $\leq1$ \\
 $\Delta_L>0$, $\Pi_L<0$ & $\leq1$ \\
 Otherwise & $=0$\\
 \hline
 \hline
\end{tabular}
\caption{The `Left' intersection number $n_i^L$ for critical points $\phi_0$ and $\phi_\pm$ when the action is ${\cal PT}$-symmetric.}
\label{tab:pt-symmetry-phipm-left}
\end{table}
\eec

For the case where $a > 0$, we split the intersection number into two parts. The number of times the thimble and anti-thimble intersect the part of the integration cycle where $x < 0$ is called $n_i^L$, and the number of times the thimble and anti-thimble intersect the part of the integration cycle where $x > 0$ is called $n_i^R$. The total intersection number thus is $n_i = n_i^L + n_i^R$. For $x < 0$, since the associated (anti-)thimble equation is a quadratic in $x$, we define the discriminant, product of roots, and sum of roots as
\beq
\Delta_L = h^2 - 4a \rho_i, ~~ \Pi_L = \frac{\rho_i}{a}, ~~ \Sigma_L = -\frac{h}{2a}. 
\eeq

We obtain similar expressions for $x > 0$
\beq
\Delta_R = h^2 + 4a\rho_i, ~~ \Pi_R = -\frac{\rho_i}{a}, ~~ \Sigma_R = \frac{h}{2a}. 
\eeq

Here we look for negative real solutions for the Eq. \eqref{eq:left-thimbles} and positive real solutions for the Eq. \eqref{eq:right-thimbles}. Standard analysis of quadratic equations gives us the conditional expressions in Tables \ref{tab:pt-symmetry-phipm-right} and \ref{tab:pt-symmetry-phipm-right}.\\

\bec
\begin{table}[!htp] 
\begin{tabular}{ c  c }
\hline
\hline 
Condition & Intersection \\
 & number \\
 \hline
 $\Delta_R>0$, $\Pi_R>0$, $\Sigma_R>0$ & $\leq2$ \\
 $\Delta_R=0$, $\Sigma_R>0$ & $\leq1$ \\
 $\Delta_R>0$, $\Pi_R<0$ & $\leq1$ \\
 Otherwise & $=0$\\
 \hline
 \hline
\end{tabular}
\caption{The `Right' intersection number $n_i^R$ for critical points $\phi_0$ and $\phi_\pm$ when the action is ${\cal PT}$-symmetric.}
\label{tab:pt-symmetry-phipm-right}
\end{table}
\eec

Combining these conditions is slightly nontrivial since there are cases where two conditional expressions cannot be satisfied simultaneously. (For example, $\Sigma_L > 0$ and $\Sigma_R > 0$ is not simultaneously possible.) Having taken care of such situations, we arrive at the results in Table \ref{tab:pt-symmetry-phipm}.

\bibliographystyle{utphys.bst}
\bibliography{references}

\end{document}